\documentstyle[11pt,aaspp4]{article}

\received{}
\journalid{}{}
\articleid{}{}

\newcommand{\ts}{\thinspace}

\slugcomment{}

\begin{document}

\title{OPTICAL SPECTROSCOPY OF THE {\em IRAS} 1-Jy SAMPLE \break 
OF ULTRALUMINOUS INFRARED GALAXIES}

\author{Sylvain Veilleux\altaffilmark{1,2}, D.-C. Kim\altaffilmark{2,3}
and D. B. Sanders\altaffilmark{4}}

\altaffiltext{1}{Department of Astronomy, University of Maryland, 
College Park, MD 20742; E-mail: veilleux@astro.umd.edu}
\altaffiltext{2}{Visiting observers at the Kitt Peak National Observatory of 
the National Optical Astronomy Observatories, operated by AURA, Inc., 
for the National Science Foundation.}
\altaffiltext{3}{Infrared Processing and Analysis Center, California
Institute of Technology, Pasadena, CA 91125; E-mail:
kim@ipac.caltech.edu}
\altaffiltext{4}{Institute for Astronomy, University of Hawaii, 2680
Woodlawn Drive, Honolulu, HI 96822; E-mail: sanders@ifa.hawaii.edu}

\begin{abstract}
This paper discusses the optical spectroscopic properties of the {\em
IRAS} 1-Jy sample ({\em f}$_{60} > 1$ Jy) of ultraluminous infrared
galaxies (ULIGs: {\em L}$_{\rm ir}$ $>$ 10$^{12}$ {\em L}$_\odot$;
{\em H}$_{\rm o}$ = 75 km s$^{-1}$ Mpc$^{-1}$ and {\em q}$_{\rm o}$ =
0).  One hundred and eight of the 118 1-Jy ULIGs have been observed at
$\Delta \lambda = 8.3${\ts}\AA~resolution over the wavelength range
$\sim$ 4500{\ts}\AA~--~8900{\ts}\AA.  These data are combined with
large, previously published sets of optical spectroscopic data of
lower luminosity infrared galaxies to look for systematic trends with
infrared luminosity over the luminosity range $L_{\rm ir} \approx
10^{10.5}-10^{13}\ L_\odot$.  As found in previous studies, the
fraction of Seyfert galaxies among luminous infrared galaxies
increases abruptly above $L_{\rm ir} \approx 10^{12.3}\ L_\odot$ ---
about 50{\ts}\% of the galaxies with $L_{\rm ir} > 10^{12.3}\ L_\odot$
present Seyfert characteristics. Many of the optical and infrared
spectroscopic properties of the Seyfert galaxies are consistent with
the presence of a genuine active galactic nucleus (AGN).  About
30{\ts}\% of these galaxies are Seyfert 1s with broad-line regions
similar to those of optical quasars.  Published near-infrared
spectroscopy also suggests that many of the Seyfert 2 galaxies
(especially those with warm {\em IRAS} 25-to-60 $\mu$m colors) are in
fact obscured Seyfert 1 galaxies with broad ($\ga$ 2,000 km s$^{-1}$)
recombination lines at 2 $\mu$m, where dust obscuration is less
important.  The percentage of Seyfert 1 ULIGs increases with infrared
luminosity, contrary to the predictions of the standard unification model
for Seyfert galaxies. Comparisons of the broad-line luminosities of
optical and obscured Seyfert 1 ULIGs with those of optically selected
quasars of comparable bolometric luminosity suggest that the dominant
energy source in most of these ULIGs is the same as in optical
quasars, namely mass accretion onto a supermassive black hole, rather
than a starburst. These results are consistent with recently published
{\em ISO}, {\em ASCA}, and VLBI data.

On the other hand, there is no unambiguous optical or near-infrared
spectroscopic evidence for an AGN in ULIGs optically classified as
H{\ts}II-region galaxies ($\sim${\ts}30{\ts}\% of the whole sample) or
as LINERs ($\sim${\ts}40{\ts}\%).  The apparent lack of an
energetically important AGN in these objects supports the results from
recent mid-infrared spectroscopy with ISO. Photoionization by hot
stars from a recent starburst appears to be the dominant source of
ionization in the objects with H{\ts}II region-like spectra, while
both hot stars and shocks may contribute to the ionization in ULIGs
with LINER-like spectra.  The weaker H$\beta$ and Mg~Ib stellar
absorption features, larger H$\alpha$ emission equivalent widths and
bluer optical continuum colors in objects of higher infrared
luminosities suggests that the starburst took place more recently
($\la$ few $\times$ 10$^7$ yrs) and/or is more important
($\sim${\ts}10{\ts}\% of the galaxy mass) in ULIGs than in their lower
luminosity counterparts.

As found in optically-selected starbursts, the emission-line gas in
ULIGs is dustier than the stellar population which is producing the
optical continuum. The color excess derived from the Balmer line ratio
does not significantly depend on the infrared luminosity, optical
spectral type, or {\em IRAS} 25-to-60 $\mu$m color of the luminous
infrared galaxies. These results suggest that the optical method used
to determine the color excess in infrared galaxies underestimates the
amount of dust in the dustier objects.
\end{abstract}


\clearpage

\section{Introduction}

The nature of the dominant energy source in ultraluminous infrared
galaxies (ULIGs: $L_{\rm ir} > 10^{12}\ L_\odot$)\footnote{$L_{\rm ir}
\equiv L (8-1000 \micron)$, computed from the observed infrared fluxes
in all four {\em IRAS} bands according to the prescription outlined in
Perault (1987)} is the object of a vigorous debate (see, e.g., review
by Sanders \& Mirabel 1996). Results from recent studies at
mid-infrared (e.g., Lutz et al. 1996, 1998; Genzel et al. 1998) and
radio wavelengths (Lonsdale et al. 1998; Smith et al. 1998a,b) suggest
that several ULIGs are powered predominantly by hot stars rather than
by an active galactic nucleus (AGN).  However, the great majority of
the infrared galaxies in these samples are of relatively modest
luminosity, with only a few having $L_{\rm ir} > 2 \times 10^{12}\
L_\odot$. This distinction may be important as there is growing
evidence from optical/near-infrared spectroscopy that the frequency of
occurence of AGN among luminous infrared galaxies (LIGs: {\em L}$_{\rm
ir}$ $\ga$ 10$^{11}$ {\em L}$_\odot$) increases with increasing
infrared luminosity.  Veilleux et al. (1995; hereafter VKSMS) carried
out a sensitive optical spectroscopic survey of a sample of 200 LIGs
and classified the nuclear spectra of these galaxies using a large
number of optical line-ratio diagnostics corrected for the underlying
stellar absorption features.  VKSMS found that 7 of the 21 ULIGs in
their sample showed Seyfert characteristics. In contrast, only 19 of
the 161 objects with lower luminosities ($L_{\rm ir} < 10^{12}\
L_\odot$) were optically classified as Seyferts.  More recently, Kim,
Veilleux, \& Sanders (1998; hereafter KVS) extended this type of
analysis to a subset of 45 ULIGs from the {\em IRAS} 1-Jy sample (Kim
1995; Kim \& Sanders 1998). The fraction of Seyfert 1 and 2 galaxies
among LIGs was found to increase dramatically above $L_{\rm ir}
\approx 10^{12.3}\ L_\odot$ with more than 50{\ts}\% of the galaxies
with $L_{\rm ir} > 10^{12.3}\ L_\odot$ having Seyfert
characteristics. Near-infrared spectroscopy of a representative subset
of ULIGs from the 1-Jy sample appears to confirm the presence of an
energetically important AGN in many of these Seyfert galaxies
(Veilleux, Sanders, \& Kim 1997, 1999b; hereafter VSK97 and VSK99,
respectively).

It is clearly important to verify the results of KVS using the entire
1-Jy sample. This sample provides a complete list of the brightest
ULIGs with F[60~$\mu$m] $>$ 1 Jy which is not biased toward `warm'
quasar-like objects. The `1-Jy' sample contains 118 objects with $z$ =
0.02 -- 0.27 and log~[L$_{\rm ir}$/L$_\odot$] = 12.00 -- 12.90.  The
infrared luminosities of these objects therefore truly overlap with
the bolometric luminosities of optical quasars. The present paper
discusses the optical spectra of 63 of the 73 ULIGs from the 1-Jy
sample that were not observed by KVS, and combine these results with
our previously published optical and near-infrared data to look for
systematic trends with infrared luminosity among galaxies with $L_{\rm
ir} \approx 10^{10.5}-10^{13}\ L_\odot$.

The structure of this paper is as follows. Section 2 discusses the
procedures used to obtain and reduce the new data.  The results
derived from these data are described and compared with those of
previous optical spectroscopic studies in \S 3.  In \S 4, the
implications of our results on the nature of the dominant energy
source in these objects are discussed in the context of recent
investigations at longer and shorter wavelengths.  The main
conclusions are summarized in \S 5.  We use $H_{\rm o} =
75$~km~s$^{-1}$~Mpc$^{-1}$ and $q_{\rm o} = 0$ throughout this paper.


\section{Observations and Data Reduction}

The new spectroscopic data were obtained with the Gold Cam
Spectrograph on the Kitt Peak 2.1-meter telescope. Table 1 lists the
dates of the observations, grating used, spectral coverage,
resolution, and seeing during the observations.  The exposure times
range from 300 sec to 3,600 sec depending on the $R$ magnitude of each
object (cf.~Kim 1995).  In all cases, a slit with a width of 2\arcsec\
was used and the slit was positioned in the N-S direction ($PA =
0^\circ$).  Most of the observations were made under photometric
conditions.  Those that were not are indicated with a dagger ($\dag$)
in Tables 3 -- 5.  The moderate seeing prevented us from extracting
useful spatial information along the spectrograph slit.  The results
derived from the KPNO data therefore only refer to the nuclear
regions. For consistency with VKSMS and KVS, the KPNO data were
reduced using the same techniques as those used in these earlier
papers. Once again, aperture-related effects were minimized using a
window for the extraction of the nuclear spectrum that varied
according to the redshift of each object so that it corresponds to a
constant linear scale (total diameter) of 4 kpc for most of the
galaxies, with the exceptions of three distant galaxies with $z$ $>$
0.2 (F00397--1312, F12032+1707, and F23499+2423)\footnote{Object names
that begin with `F' are sources identified in the {\em IRAS} Faint
Source Catalog, Version 2 (FSC: Moshir et al. 1992)} for which an
8-kpc window was used.
The spectra of the flux standard stars were used to remove the
absorption bands near 6870{\ts}\AA~and 7620{\ts}\AA~produced by
atmospheric O$_2$ (the B- and A-bands, respectively).


\section{Results}
 
The final calibrated spectra obtained at KPNO are plotted in Figure 1.
Only 10 of the 118 sources in the 1-Jy sample could not be observed
with the KPNO 2.1-meter telescope because of low declination and/or
lack of time.  Refer to Figure 1 in KVS to view the calibrated spectra
of the 45 objects observed using the Mauna Kea facilities.  Tables 2,
3 and 4 summarize the results of our analysis on the new KPNO data and
combine them with those of KVS and VKSMS.  The methods used to derive
the various quantities listed in these tables are described in detail
in Kim et al. (1995) and KVS and are summarized below.

The line fluxes were measured using two methods. First, the standard
plotting package in IRAF (``splot'') was used to measure the flux of
isolated emission lines. To deal with blended lines (e.g., H$\alpha$ +
[N~II] $\lambda\lambda$6548, 6583 and [S~II] $\lambda\lambda$6716,
6731) and emission lines affected by stellar absorption features
(H$\beta$ and H$\alpha$), we used ``specfit'', an interactive IRAF
procedure kindly provided by Gerard A. Kriss.  This routine can fit a
wide variety of emission-line, absorption-line, and continuum models
to the observed spectrum. The input parameters for the fit were
determined through ``splot'' in IRAF. We chose to fit the continuum with
a simple first-order polynomial, and the emission and absorption lines
with Gaussian profiles. The actual fitting was done via a chi-square
minimization using a simplex algorithm.  The output parameters were
the flux levels and slopes of the underlying continuum emission, the
fluxes, centroids, and widths (FWHM) of the emission lines, and
equivalent widths, centroids, and widths (FWHM) of the absorption
lines.

Uncertainties in the H$\alpha$ fluxes are typically 25 --
50\%.  The other emission line fluxes listed in Table 3 are normalized
to the observed H$\alpha$ flux. The uncertainty in these relative flux
ratios is typically 5 -- 10\% .  Colons (:) and double colons (::)
indicate values with relative uncertainties of about 25\% and 50\%,
respectively. The adopted H$\beta$ flux, listed in column (5), is the
weighted average of the values obtained from the ``splot'' and ``specfit''
methods, with weights estimated from the relative uncertainties in the
two types of measurements.

The absorption line strengths, line widths, and continuum measurements
are presented in Table 4.  Column (2) of Table 4 lists the equivalent
widths of H$\beta$ derived from the fitting method. These measurements
are rather uncertain ($\sim$50\%) because of the generally strong
H$\beta$ emission line affecting the profile of this feature.  Columns
(3) and (4) list the equivalent widths of the Mg~Ib $\lambda$5174 and
Na~ID $\lambda$5896 absorption lines, respectively. Note that the
Na~ID feature is blended with the emission line of He I at
$\lambda$5876 in some galaxies.  Column (5) of Table 4 lists the
values of the line widths (FWHM) of [O~III] $\lambda$5007. This line
was selected for line width measurements because it is strong in most
ULIGs and free of any nearby emission or absorption lines (in contrast
to H$\alpha$).  The line widths listed in Table 4 have been corrected
for the finite instrument resolution of the data ($\sim${\ts}8.3\AA)
using the quadrature method. This method assumes that the intrinsic
and instrument profiles are Gaussian and gives corrected widths that
are too large for profiles which are more peaky than Gaussians (e.g.,
the emission-line profiles in optically-selected AGN -- Whittle 1985;
Veilleux 1991). For this reason, the [O~III] $\lambda$5007 line widths
should be treated with caution. Uncertainties on these measurements
are about $\pm$ 100 -- 200 km s$^{-1}$ or more. Corrected line widths
below 100 km s$^{-1}$ are flagged with a colon in Table 4 to emphasize
the large uncertainties on these measurements. Finally, the intensity
levels of the continuum near H$\beta$ and H$\alpha$ are listed in
columns (6) and (7) as C4861 and C6563, respectively.

\subsection{Spectral Classification}

The results from the spectral analysis of the {\em IRAS} 1-Jy sample
are quantitatively similar to those derived by KVS on the subsample of
45 objects.  The emission-line ratio measurements used for this
analysis are listed in Table 5 and are plotted in Figure 2.  As
described in the previous section, these line ratios were corrected
for the underlying Balmer absorption features. For consistency with
KVS and VKSMS, the same dereddening and spectral classification
techniques were used in the analysis. The line ratios were corrected
for reddening using the values of {\em E(B--V)} determined from the
H$\alpha$/H$\beta$ ratio and the Whitford reddening curve as
parameterized by Miller \& Mathews (1972; see next section). The
boundaries of Veilleux \& Osterbrock (1987) were used to classify each
object as H~II or AGN-like galaxies.  These boundaries are based on
large data sets on optically selected galaxies and the predictions of
photoionization models by power-law spectra and by hot stars. A
distinction was made among AGN-like galaxies between the objects of
high ([O~III] $\lambda$5007/H$\beta \ge$ 3) and low ([O~III]
$\lambda$5007/H$\beta \leq$ 3) excitation.  The first group represents
the ``classic'' Seyfert 2 galaxies while galaxies in the second group
were classified as LINERs (``low-ionization nuclear emitting
regions''). This LINER definition differs from the original definition
of Heckman (1980) which partly relies on the strength of the [O~II]
$\lambda$3727 line (this lines is outside the wavelength range of our
spectra). However, results from spectroscopic studies of optically
selected objects (e.g., Veilleux \& Osterbrock 1987) suggest that the
two ways of defining LINERs are often (close to 95{\ts}\% of the time)
equivalent.  Finally, galaxies with Fe II multiplets at
$\lambda\lambda$5100--5560 and very broad ($\Delta V_{\rm FWHM}
\gtrsim 2,000$ km s$^{-1}$) H~I Balmer and He~I $\lambda$5876 emission
lines were classified as Seyfert 1s.

Following VKSMS and KVS, ULIGs with double nuclei were assigned the
more Seyfert-like spectral type of the two nuclei: galaxy pairs with
spectral types H~II -- LINER, H~II -- Seyfert 2, H~II -- Seyfert 1,
LINER -- Seyfert 2, LINER -- Seyfert 1, and Seyfert 1 -- Seyfert 2
were classified as LINER, Seyfert 2, Seyfert 1, Seyfert 2, Seyfert 1,
and Seyfert 1, respectively. These widely separated interacting
systems are more common at lower luminosities (VKSMS; Kim 1995; Surace
1998; Veilleux, Kim, \& Sanders 1999a).  Consequently, this
classification method of the double-nucleus systems is conservative in
the sense that it overestimates the fraction of AGN among the lower
luminosity objects and thus cannot explain the trends with infrared
luminosity discussed below.

Out of the 108 ULIGs that make up our spectroscopic sample, 30{\ts}\%
(32/108) were found to have spectra characteristic of photoionization
by hot stars (H~II region-like).  AGN-like emission lines were
observed in 70{\ts}\% (76/108) of the total sample including 9{\ts}\%
(10/108) Seyfert 1s, 21{\ts}\% (23/108) Seyfert 2s, and 40{\ts}\%
(43/108) LINER-like objects.  These results were combined with the
measurements obtained by VKSMS for the LIGs from the Bright Galaxy
Survey (BGS; Soifer et al. 1987, 1989) to search for systematic
variations of the spectral types with infrared luminosity. For this
exercise, ULIGs were further divided into two luminosity bins:
$10^{12}\ L_\odot \leq L_{\rm ir} < 10^{12.3}\ L_\odot$ and $L_{\rm
ir} \geq 10^{12.3}\ L_\odot$.  A summary of this analysis is given in
Table 6 and plotted in Figure 3.

This figure confirms the tendency noted by VKSMS and KVS for the more
luminous objects to be more Seyfert-like.  As noted by KVS, only one
Seyfert 1 (NGC~7469) has $L_{\rm ir} < 10^{12}\ L_\odot$, whereas
48{\ts}\% (15/31) of the galaxies with $L_{\rm ir} \geq 10^{12.3}\
L_\odot$ present Seyfert characteristics.  The percentage of Seyfert
1s relative to Seyfert 2s increases with infrared luminosity (from
$\sim$ 0\% among LIGs up to $\sim$ 50\% among ULIGs with $L_{\rm ir} >
10^{12.3}\ L_\odot$). This result is not compatible with the
predictions of the standard unification model of Seyfert galaxies
which purports that Seyfert 1s and 2s are basically the same kind of
objects observed from different angles. The far-infrared emission in
these objects is believed to be emitted more or less isotropically and
therefore does not depend on viewing angle (e.g., Mulchaey et
al. 1994).  The trend with infrared luminosity in our data can be
explained if the covering factor of the obscuring material (opening
angle of torus?)  in Seyfert ULIGs decreases with increasing infrared
luminosity.

LINER-like galaxies constitute a large fraction (30--40{\ts}\%) of the
total sample regardless of $L_{\rm ir}$.  The LIGs classified as H~II
galaxies generally have low-excitation lines ([O~III]
$\lambda$5007/H$\beta$ $<$ 3; Fig. 2) and no detectable Wolf-Rayet
emission features (e.g., combinations of He~II $\lambda$4686, He~I
$\lambda$5876, C~III $\lambda$5696, N~IV $\lambda$5737, N~III
$\lambda\lambda$4634, 4640, and 4642), confirming previous results
(KVS; VKSMS; Leech et al. 1989; Armus et al. 1989; Allen et al. 1991;
Ashby, Houck, \& Hacking 1992; Ashby et al. 1995) and strongly arguing
{\em against} a very recent burst ($<<$ 10$^7$ yr) of star formation
in these objects (cf.~\S 3.5; Evans \& Dopita 1985; McCall, Rybski, \&
Shields 1985; Allen et al. 1991).  Therefore, there is no evidence at
optical wavelengths for the type of extremely young (5 $\times$ 10$^6$
yrs) starbursts purported by Downes \& Solomon (1998) to exist in
Arp~220 and other ULIGs.  The implications of this spectral
classification are discussed in \S 4.

\subsection{Reddening}

Following KVS, the reddening in our sample galaxies was estimated
using the emission-line H$\alpha$/H$\beta$ ratios corrected for the
underlying stellar absorption features.  An intrinsic
H$\alpha$/H$\beta$ ratio of 2.85 was adopted for H II region-like
galaxies (Case B Balmer recombination decrement for T = 10$^4$ K and
N$_e$ = 10$^4$ cm$^{-3}$) and 3.10 for Seyferts and LINERs (e.g.,
Halpern \& Steiner 1983 and Gaskell \& Ferland 1984).  For consistency
with the studies of VKSMS and KVS, the Whitford reddening curve as
parameterized by Miller \& Mathews (1972) was used.  The reddenings
derived from this method are listed as color excesses, {\em E(B--V)},
in the third column of Table 5.  Since all the objects in the 1-Jy
sample lie at $|b| > 30 ^\circ$, no correction was made for Galactic
reddening.  Note that this method assumes that the obscuration is due
to a foreground screen of dust and underestimates by a factor
[exp($\tau$)--1]/$\tau$ the actual amount of extinction if the
line-emitting gas and dust are spatially mixed.  Moreover, the adopted
procedure does not take into account possible differences between
Galactic and extragalactic reddening curves (e.g., Calzetti, Kinney,
\& Storchi-Bergmann 1994). However, these differences will be small at
optical wavelengths.

The distribution of {\em E(B--V)} for the ULIGs of our sample is shown
in Figure 4.  The results are quantitatively similar to those of
KVS. Not surprisingly, the color excesses measured in the ULIGs of our
sample are considerably larger than those measured in
optically-selected Seyfert and starburst galaxies (Dahari \&
De~Robertis 1988) and in extragalactic H~II regions (Kennicutt, Keel,
\& Blaha 1989).  These results confirm the importance of dust in
infrared-selected galaxies. The median {\em E(B--V)} are 0.80, 1.11
and 1.21 for the H~II galaxies (30 objects), LINERs (45) and Seyfert 2
galaxies (22), respectively.  Kolmogorov-Smirnov (K-S) tests indicate
that the differences between the various spectral types are not
significant. These color excesses are similar to those obtained by
VKSMS in the lower-luminosity galaxies of the BGS [{\em E(B--V)} =
1.05, 1.24, and 1.07 for H~II galaxies, LINERs, and Seyfert 2
galaxies, respectively]. However, VKSMS found that the color excesses
of the LINERs were significantly larger than those of the H~II and
Seyfert 2 galaxies; this difference is not observed among the 1-Jy ULIGs.

The correlation reported by VKSMS between {\em E(B--V)} and the equivalent
width of the Na{\ts}ID absorption feature, {\em EW}(Na{\ts}ID),   in 
H~II galaxies is also present, but at a weaker level,
among our high-luminosity objects.  The probability, P[null], that
this correlation is fortuitous is 0.04, 0.01, and 0.001 among the
H~II galaxies, LINERs, and Seyfert 2 galaxies of the 1-Jy sample. An
important fraction of the Na{\ts}ID feature is therefore of interstellar
origin. A similar correlation is observed between {\em E(B--V)} and
the observed continuum colors of H~II galaxies (P[null] = 3 $\times$
10$^{-5}$) and LINERs (P[null] = 1 $\times$ 10$^{-4}$), but not among
Seyfert 2 galaxies (P[null] = 0.16).  As first reported by VKSMS,
these results suggest that the scatter in the continuum colors of the
Seyfert LIGs is predominantly intrinsic rather than caused by
variations in the amount of reddening from one object to the other
(cf. \S 3.5 for a more detailed discussion of continuum colors).

The {\em IRAS} flux density ratio ${\em f}_{25}/{\em f}_{60}$ is a
well-known indicator of Seyfert activity in infrared galaxies (e.g.,
deGrijp et al 1985; Miley, Neugebauer \& Soifer 1985). Galaxies with
``warm'' {\em IRAS} 25-to-60 $\mu$m colors are generally believed to
be less affected by dust obscuration than the ``cooler'' objects
(e.g., Sanders et al 1988a,b; Surace et al. 1998; VSK97, VSK99). One
should therefore expect to see a positive correlation between the {\em
IRAS} 25-to-60 $\mu$m color and the color excess in our sample;
however, none is detected (Fig. 5).  This result suggests that the
color excess derived from the optical emission lines is not always a
good indicator of the total column of material towards the nucleus of
these objects. This problem is more severe in the cooler, dustier
objects because the optical method saturates beyond $\tau_{\rm V} \sim
5$ or, equivalently, {\em E(B--V) $\sim${\ts}2}. Extinction
measurements at longer wavelengths are more reliable (e.g., VSK97,
VSK99, Genzel et al. 1998).

\subsection{Line Widths}

Figure 6 shows the distribution of the [O~III] line widths for each
spectral type. The median line widths of the H II galaxies and LINERs
are comparable (200 km s$^{-1}$ and 320 km s$^{-1}$) whereas the
[O~III] line widths of Seyfert 1 and 2 galaxies are larger (1,120 and
610 km s$^{-1}$, respectively). K-S tests confirm these
differences. The probability that the line widths of H~II galaxies and
LINERs are drawn from the same population is 0.70, but it is less than
0.001 when comparing the line widths of H~II and Seyfert galaxies, or
when comparing the LINERs and Seyfert galaxies.

Figure 7 presents the distribution of [O~III] line widths of the
combined sample as a function of infrared luminosity and as a function
of the {\em IRAS} flux density ratio ${\em f}_{25}/{\em f}_{60}$.  The
new data strengthen the conclusions of VKSMS and KVS: there is no
obvious correlation between line width and infrared luminosity or {\em
IRAS} color, but nearly all of the objects with line widths
larger than 600 km s$^{-1}$ have $L_{\rm ir} \gtrsim 10^{11}\ L_\odot$
(the only exception is the H~II galaxy NGC~3597 which has log~[$L_{\rm
ir}$/L$_\odot$] = 10.91). Objects with the most extreme profiles
($\Delta V_{\rm FWHM} \ga 1200$ km s$^{-1}$) all have $L_{\rm ir} \ga
10^{ 12}\ L_\odot$, optical Seyfert characteristics, and warm {\em
IRAS} colors (${\em f}_{25}/{\em f}_{60}$ $\ga$ 0.12).  These results
suggest that the nuclear activity in the more powerful Seyfert ULIGs
contributes to the line broadening.  Non-gravitational processes
associated with AGN-driven outflows have been proposed by VKSMS and
KVS to explain the broad and complex line profiles in the nuclear and
circumnuclear regions of some LIGs.

\subsection{Stellar Absorption Features}

In Figures 8 and 9, the distributions of the equivalent widths of
H$\beta$ and Mg~Ib are presented as a function of spectral type.
These distributions are similar to those obtained by KVS in their
subset of the 1-Jy sample. The conclusions derived by KVS on the
stellar absorption features are therefore generally confirmed by the
analysis of the entire 1-Jy sample.  The median {\em EW}(H$\beta$) for
the 1-Jy ULIGs (including those with undetected H$\beta$) are
0.8{\ts}\AA, 0.6{\ts}\AA, 0.0{\ts}\AA\, and 0.0{\ts}\AA\ for the H~II
galaxies (30 objects), LINERs (43), Seyfert 2 galaxies (22), and
Seyfert 1 galaxies (9), respectively. K-S tests indicate that the
equivalent widths of Seyfert 1 and 2 galaxies are significantly
smaller than those of LINERs and H~II galaxies (P[null] $\la$ 0.01).
The equivalent widths of the 1-Jy ULIGs taken as a whole are also
significantly smaller than those measured by VKSMS in the BGS LIGs
(P[null] = 4 $\times$ 10$^{-9}$ when comparing both samples, and
P[null] = 0.002 and 0.01 for LINERs and H~II galaxies, respectively.
There are too few Seyfert galaxies among the BGS LIGs to carry out
this analysis).  The intermediate-age (10$^8$--10$^9$ yr) population
of stars that is responsible for the H$\beta$ absorption feature is
therefore less prominent in our sample of ULIGs, possibly because the
H$\beta$ feature is diluted by a hot, featureless continuum from
young stars (particularly in H~II galaxies) or from an AGN (in the
Seyfert galaxies).

The strength of the Mg~Ib absorption feature provides additional
constraints on the underlying stellar population in ULIGs.  As found
by KVS, the {\em EW}(Mg Ib) for our sample of ULIGs [median values of
0.0{\ts}\AA, 0.1{\ts}\AA, 0.2{\ts}\AA, and 0.0{\ts}\AA~for H II
galaxies (30 objects), LINERs (43), Seyfert 2 galaxies (22), and
Seyfert 1 galaxies (9), respectively] are considerably smaller that
those measured in non-active spiral galaxies [{\em EW}(Mg~Ib) =
3.5--5.0{\ts}\AA: Keel 1983; Stauffer 1982; Heckman, Balick, \& Crane
1980]. Moreover, {\em contrary} to KVS findings, the {\em EW}(Mg Ib)
for our sample of ULIGs are also significantly smaller than those
measured in the BGS objects of VKSMS (1.12{\ts}\AA, 1.49{\ts}\AA, and
1.17{\ts}\AA~for H~II galaxies, LINERs, and Seyfert 2 galaxies,
respectively).  The very weak H$\beta$ and Mg~Ib features in 1-Jy
ULIGs can be explained if the age of the {\em dominant} stellar
population in these objects is less than a few $\times$ 10$^7$ yrs and
$\sim${\ts}10{\ts}\% of the galaxy mass is taking part in a burst of
star formation (Bica, Alloin, \& Schmidt 1990). Alternatively, an AGN
may account for $\sim${\ts}50 -- 75{\ts}\% of the optical continuum in
some ULIGs (e.g., Seyfert 1 and 2 galaxies) and weaken the spectral
signatures of the underlying old stellar population. The analysis of
the 1-Jy sample indicates that the young stellar population and/or AGN
is more preeminent in ULIGs than in the lower luminosity objects.

\subsection{Continuum Colors}

The observed continuum colors of the 1-Jy ULIGs are presented in
Figure 10 as a function of spectral type. Seyfert 1s have continuum
colors (median C6563/C4861 of 0.70) which are significant bluer than
all other classes of objects (median C6563/C4861 = 0.98, 1.01, and
1.15 for H~II galaxies, LINERs, and Seyfert 2 galaxies, respectively).
This result is consistent with the presence in Seyfert 1s of an AGN
emitting a strong energetic continuum. Figure 10 also shows that the
optical continuum of ULIGs is on average bluer than that of the BGS
LIGs (P[null] = 2 $\times$ 10$^{-7}$ when considering all
objects). This is especially evident among the LINERs (P[null] = 4
$\times$ 10$^{-5}$) and the H{\ts}II galaxies (P[null] = 2 $\times$
10$^{-3}$). These results confirm the presence of a strong blue
featureless continuum from a young stellar population and/or AGN in
the high-luminosity objects (\S 3.4). 

The expected continuum color, C6563/C4861, of galaxies where
$\sim${\ts}10{\ts}\% of the galaxy mass has taken part in a burst of
star formation less than a few 10$^7$ yrs ago is $\sim${\ts}0.90
(e.g., Bica, Alloin, \& Schmidt 1990). Yet, the continuum colors
measured in ULIGs generally are redder than this value.  This is a
sign that the continuous emission from these objects is probably
affected by dust obscuration.  However, if the observed continuum
colors are dereddened using the amount of dust derived from the
emission-line spectrum (\S 3.2), a strong {\em negative} correlation
becomes apparent between the dereddened continuum colors and
dereddened H$\alpha$ luminosities (P[null] = 10$^{-14}$ for the whole
sample; in contrast, P[null] = 0.1 when the {\em observed} continuum
colors are considered).  As mentioned in KVS, this result strongly
suggests that this dereddening method overcorrects for reddening in
the continuum (cf. also \S 3.7).  The emission-line gas near hot stars
or the AGN are therefore dustier than the relatively cold stellar
population which is producing the optical continuum.  This discrepancy
is also observed among optically-selected starbursts (e.g., Calzetti
et al. 1994; Gordon, Calzetti, \& Witt 1997).  We have no easy way to
properly deredden the continuum colors of our sample galaxies. The
dereddened continuum colors will not be considered any further in the
present discussion.

\subsection{H$\alpha$ Luminosities and Equivalent Widths}

Here again, many of the conclusions based on the smaller sample of KVS
are confirmed in the complete 1-Jy sample.  The distributions of
H$\alpha$ equivalent widths, observed H$\alpha$~luminosities, and
infrared-to-H$\alpha$~luminosity ratios are plotted in Figures 11 --
13 for each spectral type.  The large H$\alpha$ luminosities and
equivalent widths and small infrared-to-H$\alpha$~luminosity ratios
among the (4) Seyfert 1 ULIGs [their median {\em EW}(H$\alpha$) in
\AA, log ({\em L}$_{{\rm H}\alpha}$/{\em L}$_\odot$), and log({\em
L}$_{\rm ir}$/{\em L}$_{{\rm H}\alpha}$) are 234, 8.6, and 3.8,
respectively] again suggest the presence of an additional source of
ionization (AGN) which is not visible or present in starburst-powered
galaxies.  LINER ULIGs appear to be slightly deficient in H$\alpha$
relative to H~II and Seyfert 2 ULIGs: the median \{{\em EW}(H$\alpha$)
in \AA, log ({\em L}$_{{\rm H}\alpha}$/{\em L}$_\odot$)\} are \{88,
7.8\}, \{50, 7.4\}, and \{84, 7.9\} for H~II galaxies (30 objects),
LINERs (43), and Seyfert 2 galaxies (22), respectively.  The observed
infrared-to-H$\alpha$~luminosity ratios in LINER ULIGs are also
somewhat larger than those of Seyfert 2 and H~II ULIGs [median
log({\em L}$_{\rm ir}$/{\em L}$_{{\rm H}\alpha}$) = 4.2, 4.7, and 4.2
for the H~II (29), LINER (38), and Seyfert 2 galaxies (21),
respectively], confirming the H$\alpha$ deficiency in LINERs.  This
deficiency persists even after correcting the emission-line fluxes for
reddening using the color excesses derived from the optical Balmer
decrements. The H$\alpha$ equivalent widths of HII and LINER ULIGs are
slightly larger than those found in LIGs of the same type (55 \AA~in
HII galaxies and 29 \AA~in LINERs). This is consistent with the
presence of a more preeminent starburst in these ULIGs (\S 3.4 and \S
3.5).

\subsection{Infrared Spectral Properties}

The infrared spectral properties of the various classes of LIGs were
investigated. We used the definitions of Dahari \& De Robertis (1988)
for the ``{\em IRAS} color indices'' and ``infrared color excess''
[$\alpha_1$ = -- log(${\em f}_{60}/{\em f}_{25}$)/log(60/25),
$\alpha_2$ = -- log(${\em f}_{100}/{\em f}_{60}$)/log(100/60), $IRCE$
= (($\alpha_1$ + 2.48)$^2$ + ($\alpha_2$ + 1.94)$^2$)$^{0.5}$].  The
infrared color excess is a measure of the deviation from colors of
non-active spiral galaxies ($\alpha_1$ = --2.48 and $\alpha_2$ =
--1.94; Sekiguchi 1987).

As described in Kim \& Sanders (1998), the {\em IRAS} 60-to-100 flux
ratios were used to pre-select the targets for the 1-Jy sample.
Consequently, the 1-Jy ULIGs have median $\alpha_2$ and $IRCE$ (--0.41
and 1.63, respectively) which are significantly larger than those of
the BGS LIGs (--0.77 and 1.19, respectively), while the median
$\alpha_1$ for ULIGs and BGS LIGs (--2.31 and --2.46, respectively)
are similar. The differences in {\em IRAS} colors between ULIGs and
normal galaxies are especially striking among Seyfert 1 galaxies
(median $\alpha_1$, $\alpha_2$, and $IRCE$ of --1.70, +0.05, and 2.16,
respectively) and Seyfert 2 galaxies (--2.27, --0.38, and 1.61).  The
infrared spectral properties of the LINER ULIGs (--2.49, --0.53, and
1.53) are indistinguishable from those of H{\ts}II ULIGs (--2.49,
--0.45, 1.59) and are closer to the parameters of normal spiral
galaxies. Therefore, there is no convincing optical or infrared
evidence for an AGN in LINER ULIGs. VKSMS and KVS argue that hot stars
and/or shocks are the most likely sources of ionization in these
objects (cf. \S 4).

Optically classified H~II galaxies in our sample show correlations
between infrared colors and H$\alpha$ equivalent widths.  H~II
galaxies with large H$\alpha$ equivalent widths tend to have small
${\em f}_{12}/{\em f}_{25}$ and large ${\em f}_{25}/{\em f}_{60}$ and
${\em f}_{60}/{\em f}_{100}$ (P[null] = 0.001, 0.002 and 4 $\times$
10$^{-9}$, respectively; Fig. 14).  Similar correlations were reported
by Mazzarella, Bothun, \& Boroson (1991) among optically-selected
starburst galaxies.  These results suggest that the most active or
youngest star-forming regions produce far-infrared emission with the
warmest dust temperatures (Mazzarella et al. 1991). The fact that
these correlations are considerably weaker among the infrared-selected
Seyfert 2 galaxies (P[null] = 0.11, 0.005 and 0.002, respectively) is
another indication that star formation may not be the only source of
energy in these objects.

As shown in Figure 15, the well-known correlation between the
optical-to-infrared luminosity ratio and the infrared luminosity or
{\em IRAS} flux ratio ${\em f}_{60}/{\em f}_{100}$ among luminous
infrared galaxies (e.g., VKSMS) disappears when the 1-Jy ULIGs are
included in the analysis.  In this figure, the optical continuum
luminosity, L(4861), is defined as P(4861) $\times$ 4861 where P(4861)
is the monochromatic power of the continuum at 4861 \AA. No reddening
correction was applied to the continuum luminosity.  Note that L(4861)
refers to the nuclear region while L$_{\rm ir}$ and ${\em f}_{60}/{\em
f}_{100}$ are integrated quantities. Integrated continuum luminosities
are not yet available for most of these galaxies. However, recent imaging
studies at infrared and millimeter wavelengths (e.g., Evans 1998;
Egami 1998; Scoville et al. 1998; Downes \& Solomon 1998; Sakamoto et
al. 1999) now strongly suggest that the active regions in ULIGs are
generally centered on the inner kpc of the galactic nuclei, with the
rest of the galaxy contributing very little to the bolometric
luminosity (the situation is different in objects of lower
luminosities).  The correlations among lower luminosity objects have
been interpreted in the past as an indication that high global dust
temperature is associated with large infrared dust emission and small
optical continuum extinction.  The tendency reported by VKSMS for the
color excess of H~II and LINER LIGs from the BGS sample to decrease at
large log~L(4861)/L$_{ir}$ is consistent with this hypothesis, but
this tendency disappears when the 1-Jy ULIGs are included in the
analysis. It is not clear at present why that is the case. An attempt
to correct L(4861)/L$_{ir}$ for reddening effects using the color
excesses determined from the H$\alpha$/H$\beta$ flux ratios fails to
bring new insight into this issue. Instead, a very strong {\em
positive} correlation between {\em E(B--V)} and L(4861)$_0$/L$_{ir}$
(P[null] = 10$^{-43}$ when considering all the objects) is created,
confirming that the reddening correction derived from the emission
lines severely overestimates the actual extinction of the optical
continuum (\S 3.5).


\section{Discussion: Nature of the Energy Source}

Most of the trends observed by KVS in their subset of 45 galaxies from
the 1-Jy sample are confirmed in the present analysis of the entire
sample. Most relevant to the question, raised in the Introduction, of
the nature of the energy source in ULIGs is the confirmation that
Seyfert 1 or 2 spectral characteristics are present in
$\sim${\ts}30{\ts}\% of all ULIGs in the 1-Jy sample, and in
$\sim${\ts}50{\ts}\% of the high-luminosity objects (log[L$_{\rm
ir}$/L$_\odot$] $>$ 12.3). The 
weaker H$\beta$ and Mg~Ib stellar absorption features, bluer continuum
colors, larger H$\alpha$ luminosities and equivalent widths, smaller
infrared-to-H$\alpha$ luminosity ratios, and warmer {\em IRAS} ${\em
f}_{25}/{\em f}_{60}$ colors in these Seyferts relative to LINER or
H~II ULIGs are all consistent with the presence of a genuine AGN in
these objects.  The absence of correlations between line widths and
line ratios (not shown here) argues against the possibility that
high-velocity shocks (e.g., Binette, Dopita, \& Tuohy 1985; Innes
1992; Dopita \& Sutherland 1995) are the dominant source of ionization
in these Seyfert 2 ULIGs. The very restrictive range of shock
conditions (e.g., V$_{\rm shock}$ = 300 -- 500 km s$^{-1}$) needed to
produce the high-excitation spectrum of Seyfert 2 galaxies from
high-velocity shocks also is inconsistent with the broad range in line
widths observed in these objects (cf. also VKSMS).

Recent near-infrared spectroscopy of 22 optically classified Seyfert 2
galaxies from the 1-Jy sample by VSK97 and VSK99 reveals the presence
of broad recombination lines with FWHM $\ga$ 2,000 km s$^{-1}$ or
strong high-ionization [Si~VI] emission in $\sim${\ts}70{\ts}\% of
these objects (especially those with warm {\em IRAS} 25-to-60 $\mu$m
colors). The absence of broad features in the profiles of the H$_2$
lines and forbidden [Fe~II]~$\lambda$1.257 suggests that the broad
emission is produced by high-density gas near the nucleus like that in
the BLR of AGN rather than by outflowing lower density material in the
circumnuclear region. The bright [Si~VI] feature detected in these
objects is also believed to be a good indicator of AGN activity since
this feature has never been detected in starburst galaxies despite
extensive searches (e.g., Marconi et al.  1994). The optical and
near-infrared results taken together, therefore suggest that the total
fraction of objects in the 1-Jy sample with signs of bonafide AGN is
at least $\sim${\ts}20 -- 25{\ts}\%, but reaches $\sim${\ts}35 --
50{\ts}\% for those objects with log[L$_{\rm ir}$/L$_\odot$] $>$ 12.3.
Nevertheless, the presence of an AGN in ULIGs does not necessarily
imply that AGN activity is the dominant energy source in these
objects. A more detailed look at the AGN in these ULIGs is needed to
answer this question.

A strong linear correlation has long been known to exist between the
continuum (or, equivalently, bolometric) luminosities of broad-line
AGN and their emission-line luminosities (e.g., Yee 1980; Shuder 1981;
Osterbrock 1989).  This correlation has often been used to argue that
the broad-line regions in AGN are photoionized by the nuclear
continuum. If this is the case, the broad-line--to--bolometric
luminosity ratio is a measure of the covering factor of the BLR (e.g.,
Osterbrock 1989). This correlation was also used by VSK97 to estimate
the importance of the AGN in powering ULIGs. In ULIGs powered uniquely
by an AGN, we expect the broad-line luminosities to fall along the
correlation for AGN. Any contribution from a starburst will increase
the bolometric luminosity of the ULIG without a corresponding increase
in the broad-line luminosity.  Starburst-dominated ULIGs are therefore
expected to fall below the ``pure-AGN'' correlation traced by the
optical quasars in a diagram of $L_{\rm H\beta}$(BLR) plotted as a
function of $L_{\rm bol}$\footnote{Note that the origin of the
far-infrared/submm emission in optical quasars is still a matter of
some debate. Some of it may be produced by a dusty circumnuclear
starburst (e.g., Rowan-Robinson 1995).  In the following discussion,
we identify the ``pure-AGN'' $L_{\rm H\beta}$(BLR) -- $L_{\rm bol}$
relation as the one traced by optical quasars. We therefore implicitly
assume that all of the far-infrared/submm emission is powered by the
central AGN (Sanders et al. 1989). The far-infrared/submm emission
generally contributes about 20\% of the total bolometric luminosities
of optical quasars. }.

In Figure 16, the dereddened {\em broad-line} H$\beta$ luminosities of
the Seyfert 1 ULIGs in our sample are plotted as a function of their
bolometric luminosities. Data on optically selected quasars and on
ULIGs with obscured BLRs (i.e. ``buried quasars'' detected at
near-infrared wavelengths by VSK97) are shown for comparison. $L_{\rm
bol}$ for the QSOs was determined using the bolometric correction
factor 11.8 (i.e. $L_{\rm bol} = 11.8 \nu_{\rm B} L_{\rm \nu}(B)$:
Elvis et al. 1994; Sanders \& Mirabel 1996), except for those few
sources that were detected by {\em IRAS}, in which case $L_{\rm bol}$
was taken from Sanders et al. (1989).  For the 15 ULIGs $L_{\rm bol}$
was taken to be $1.15 \times L_{\rm ir}$ (Kim \& Sanders 1998).  The
H$\beta$ data for the optically selected QSOs are from Yee (1980)
corrected for $H_{\rm o} = 75$~km~s$^{-1}$~Mpc$^{-1}$ and $q_{\rm o} =
0$.  The H$\beta$ luminosities tabulated in Yee (1980) include a
contribution from narrow H$\beta$ and are not corrected for
reddening. The flux from narrow H$\beta$ is often difficult to measure
in these broad-line objects, but the high-quality rest-frame optical
spectra that exist on a few of these quasars show that $L_{\rm
H\beta}$(NLR)/$L_{\rm H\beta}$(BLR) is generally much less than
$\sim${\ts}30{\ts}\% (e.g., Boroson \& Green 1992; Corbin \& Boroson
1996). This correction would lower the quasar data points in Figure
16.  The reddening correction to the broad-line luminosities would, on
the other hand, raise these data points.  The reddening towards the
BLRs of these {\em optically-selected} objects is likely to be
modest. In the end, no attempt was made to correct for narrow-line
contamination or reddening effects.  The best log-linear fit through
the quasar data points, log $L_{H\beta}$(BLR) = 1.05 log $L_{\rm bol}$
-- 3.61, is shown as a solid line in Figure 16. The broad-line
H$\beta$ luminosities of the ULIGs with optical BLRs were dereddened
using the broad-line H$\alpha$/H$\beta$ ratio and assuming an
intrinsic ratio of 3.1 (see MacAlpine 1981 for uncertainties
associated with using this dereddening method for BLRs).  The
broad-line H$\beta$ luminosities of the ULIGs with obscured BLRs were
calculated from the measured broad Pa$\alpha$ fluxes assuming Case B
recombination (except for Mrk~463E = F13536+1836 where the broad
Pa$\beta$ flux was used).  The reddening correction for these objects
was carried out using the color excesses derived from the infrared
broad-line ratios (cf.~VSK97).

Considering the various sources of uncertainties in the derivations of
the dereddened broad-line luminosities plotted in Figure 16 (overall
uncertainties of order 30{\ts}\%) and the intrinsic scatter of the
quasar data points around the ``pure'' quasar relation (of order 0.2
dex in log[L$_{\rm bol}$]), caution should be used when using this
figure to make quantitative statements about the dominant energy
source in these objects. The positions of 6 or 7 of the 10 optical
BLRs (Mrk~231 = F12540+5708, F07599+6508, F13218+0552, F13342+3932,
F15462-0450, F21219-1757 and perhaps also Mrk~1014 = F01572+0009) and
of all 5 obscured BLRs (Mrk~463E, PKS~1345+12, F20460+1925,
F23060+0505, and F23499+2423) fall close (or above) to that of optical
quasars in Figure 16. This result suggests that a substantial fraction
of the luminosity in these objects is powered by a quasar rather than
a starburst. The only broad-line ULIGs which clearly fall below the
``pure-AGN'' correlation in Figure 16 are NGC~7469, F11119+3257, and
F11598-0112. NGC~7469 has long been known to host a Seyfert 1 nucleus
and a powerful circumnuclear starburst (De~Robertis \& Pogge 1986;
Wilson et al. 1986). Recent ISO results (Genzel et al. 1998) indicate
that the starburst is indeed the dominant infrared energy source in
this object. The only other broad-line 1-Jy ULIGs in common with the
ISO sample are Mrk~231, Mrk~463E and F23060+0505. In all three cases,
the results from the ISO survey support our conclusion that the
dominant source of infrared radiation is a quasar rather than a
starburst (Genzel et al. 1998).

It is also instructive to study the mid-infrared results on the
optically classified LINERs of our sample. Five of these objects were
studied by Genzel et al. (1998; UGC 5101, F12112+0305, F14348-1447,
F15250+3609, and Arp 220 = F15327+2340). All of them are considered powered
predominantly by a starburst (Genzel et al. 1998). This result is also
confirmed when a larger sample of ULIGs is considered: if at all
present, the AGN in ULIGs with optical LINER characteristics does not
appear energetically important in the great majority of these objects
(Lutz, Veilleux, \& Genzel 1999).

Another method commonly used to estimate the importance of the AGN in
ULIGs is based on the strength of the hard (2 -- 10 keV) X-ray
emission from these objects. Nakagawa et (1998) have recently
summarized the ASCA results on ULIGs. Among our subset of broad-line
ULIGs, three objects were observed by ASCA: Mrk 231, F20460+1925, and
F07599+6508. The first two objects show clear signs of AGN activity in
the hard X-rays while the last object was not detected by ASCA.
Assuming a X-ray luminosity fraction $L$(2 -- 10 keV)/$L_{\rm FIR}$
$\sim$ 0.1 for typical optical quasars, Nakagawa et al. concluded that
the AGN contribution to the total luminosity is significant in
F20460+1925, in agreement with our optical data, but is small in
Mrk~231 and F07599+6508, in contrast to the optical results (and the
ISO results on Mrk~231).  A multiwavelength analysis of the
Palomar-Green Bright Quasars (Sanders et al. 1989) shows, however,
that the X-ray luminosity fraction is typically only about 1{\ts}\%
(and can be as small as 0.1{\ts}\% in some objects).  The AGN contribution
in Mrk~231 would be significant if Mrk 231 happens to lie in the faint
X-ray tail of the optically selected quasars. The same conclusion
applies to F07599+6508 if the actual hard X-ray flux from this object
is close to the upper limit determined by Nakagawa et
al. Consequently, the current X-ray data on the broad-line ULIGs of
our sample are not inconsistent with the presence of energetically
important AGN in many of these broad-line objects.

High-resolution radio observations of several ULIGs have provided
additional information on the nature of the energy source of ULIGs. A
good correlation seems to exist between the core radio power and
bolometric luminosity of optically-selected quasars (Lonsdale, Smith,
Lonsdale 1995). This correlation was used by Lonsdale et al. to argue
that buried quasars are capable of powering the infrared luminosities
in most ULIGs. However, recent VLBI observations by the same group
indicate that a starburst may be able to explain the observed radio
characteristics of at least a few of these objects (e.g., Arp 220 West;
Smith et al. 1998b, Lonsdale et al. 1998).  The current VLBI sample
includes only two broad-line galaxies from our sample, Mrk~231 and
NGC~7469.  The VLBI data on these two objects are difficult to explain
with a starburst alone, and therefore strongly suggest the presence of
an AGN in both objects (Smith et al 1998a).  High-resolution radio
maps of more broad-line ULIGs from our sample will be needed to
properly test our conclusions of the existence of energetically
important AGN in these objects.


\section{Summary}

The results from an optical spectroscopic study of the nuclear regions
of 108 of the 118 ULIGs from the {\em IRAS} 1-Jy sample of Kim \&
Sanders (1998) are reported. These spectra are combined with
previously published optical data on 200 LIGs from the BGS sample
(VKSMS) to examine the spectral properties of LIGs over the range
$L_{\rm ir} \approx 10^{10.5}-10^{13.0}\ L_\odot$.  The results from
this analysis confirm many of the trends already seen by KVS in a
subset of 45 of these objects. The fraction of LIGs with Seyfert
characteristics increases with increasing $L_{\rm ir}$.  For $L_{\rm
ir} > 10^{12.3}\ L_\odot$, about 48{\ts}\% of the ULIGs (15/31
objects) are classified as Seyfert galaxies. The fraction of Seyfert
1s relative to Seyfert 2s increases with infrared luminosity.  This
result is not compatible with the predictions of the standard
unification model of Seyfert galaxies unless the covering factor of
the obscuring material (opening angle of torus?)  in Seyfert ULIGs
decreases with increasing infrared luminosity.

Many of the optical and infrared spectroscopic properties of the
Seyfert galaxies point to the existence of an AGN which is not present
or visible in LINER or H~II ULIGs.  About 30{\ts}\% (10/33) of the
Seyfert galaxies in the 1-Jy sample are of type 1, presenting broad
Balmer lines and strong Fe~II emission similar to what is observed in
optically selected quasars.  Near-infrared spectroscopy (VSK97, VSK99)
often reveals broad (FWHM $>$ 2,000 km s$^{-1}$) Paschen emission
lines or strong high-ionization [Si~VI] emission in the remaining Seyferts
(especially those with warm {\em IRAS} 25-to-60 $\mu$m
colors). Seyfert ULIGs (especially those of type 1) have weaker
H$\beta$ and Mg~Ib stellar absorption features, bluer continuum
colors, larger H$\alpha$ luminosities and equivalent widths, smaller
infrared-to-H$\alpha$ luminosity ratios, and warmer {\em IRAS} ${\em
f}_{25}/{\em f}_{60}$ colors than LINER or H~II ULIGs.  The [O~III]
$\lambda$5007 line widths in the nuclei of the Seyfert galaxies are
also significantly broader on average than those measured in the H~II
and LINER ULIGs.

Comparisons between the emission-line luminosities of the optical or
near-infrared broad-line regions in the Seyfert galaxies of the 1-Jy
sample and the broad-line luminosities of optical quasars suggest that
the AGN in these ULIGs generally is energetically important. Only
three objects (NGC~7469, F11119+3257, and F11598-0112) are clearly
powered predominantly by a starburst.  The general agreement between
this optical analysis and the current ISO, X-ray, and VLBI results is
encouraging but only based on a limited number of broad-line
objects. Results from on-going multiwavelength studies of
high-luminosity ULIGs (especially those with $L_{\rm ir} \ga
10^{12.3}\ L_\odot$) will provide more stringent constraints to test
these conclusions.

The weak H$\beta$ and Mg~Ib features in ULIGs optically classified as
H~II galaxies or LINERs suggests the presence of a young ($\la$ few
$\times$ 10$^7$ yrs) stellar population comprising
$\sim${\ts}10{\ts}\% of the total galaxy mass in these objects.  While
this starburst is the likely source of ionization among H~II galaxies
and some LINERs, long-slit information from VKSMS and KVS indicates
that the LINER-like emission may also be produced through shocks
caused by the interaction of starburst-driven outflows with the
ambient material.  Comparisons of the optical results with recent
mid-infrared data obtained with ISO suggest that the main source of
energy in these infrared-selected H~II galaxies and LINERs is a
starburst rather than an AGN. The observed continuum colors and
strengths of the stellar absorption features and H$\alpha$ emission
line indicate that the starburst becomes increasingly important with
increasing infrared luminosity in both H~II galaxies and LINERs.

The large H$\alpha$/H$\beta$ ratio and red continuum colors of the
1-Jy ULIGs confirm the importance of dust in all these objects. As was
found among optically selected starbursts, the relatively cool stars
which are producing the continuous optical emission are less reddened
than the emission-line gas in ULIGs.  No significant differences are
found between the mean emission-line color excess of ULIGs and that of
{\em IRAS} galaxies of lower infrared luminosity. The color excess in
the nuclei of ULIGs does not depend strongly on the optical spectral
type or {\em IRAS} 25-to-60 $\mu$m color. The most likely explanation
for this surprising result is that the optical method used to derive
the color excess in infrared galaxies underestimates the amount of
dust in the dustier, cooler objects.

\clearpage

\acknowledgments 

S. V. and D. B. S. thank the organizers of the 1998 Ringberg meeting
where some of the issues in this paper were discussed. We are also
grateful to the editor, Dr. Greg Bothun, and the anonymous referee for
suggestions which significantly improved this paper. This research was
supported in part by JPL contract no. 961566 to the University of
Hawaii (D. B. S.). S. V. is grateful for partial support of this
research by NASA/LTSA grant NAG 56547 and Hubble fellowship
HF-1039.01-92A awarded by the Space Telescope Science Institute which
is operated by the AURA, Inc. for NASA under contract
No. NAS5--26555. This research has made use of the NASA/IPAC
Extragalactic Database (NED) which is operated by the Jet Propulsion
Laboratory, California Institute of Technology, under contract with
NASA.

\clearpage

\centerline{FIGURE CAPTIONS}

\vskip 0.3in

\figcaption[fig1.eps]{Optical spectra of the ULIGs in the KPNO sample ---
$f_\lambda$ is plotted vs $\lambda_{\rm observed}$.  The units of the
vertical axis are 10$^{-15}$ erg s$^{-1}$ cm$^{-2}$ \AA$^{-1}$ while
the wavelength scale is in{\ts}\AA.  }

\figcaption[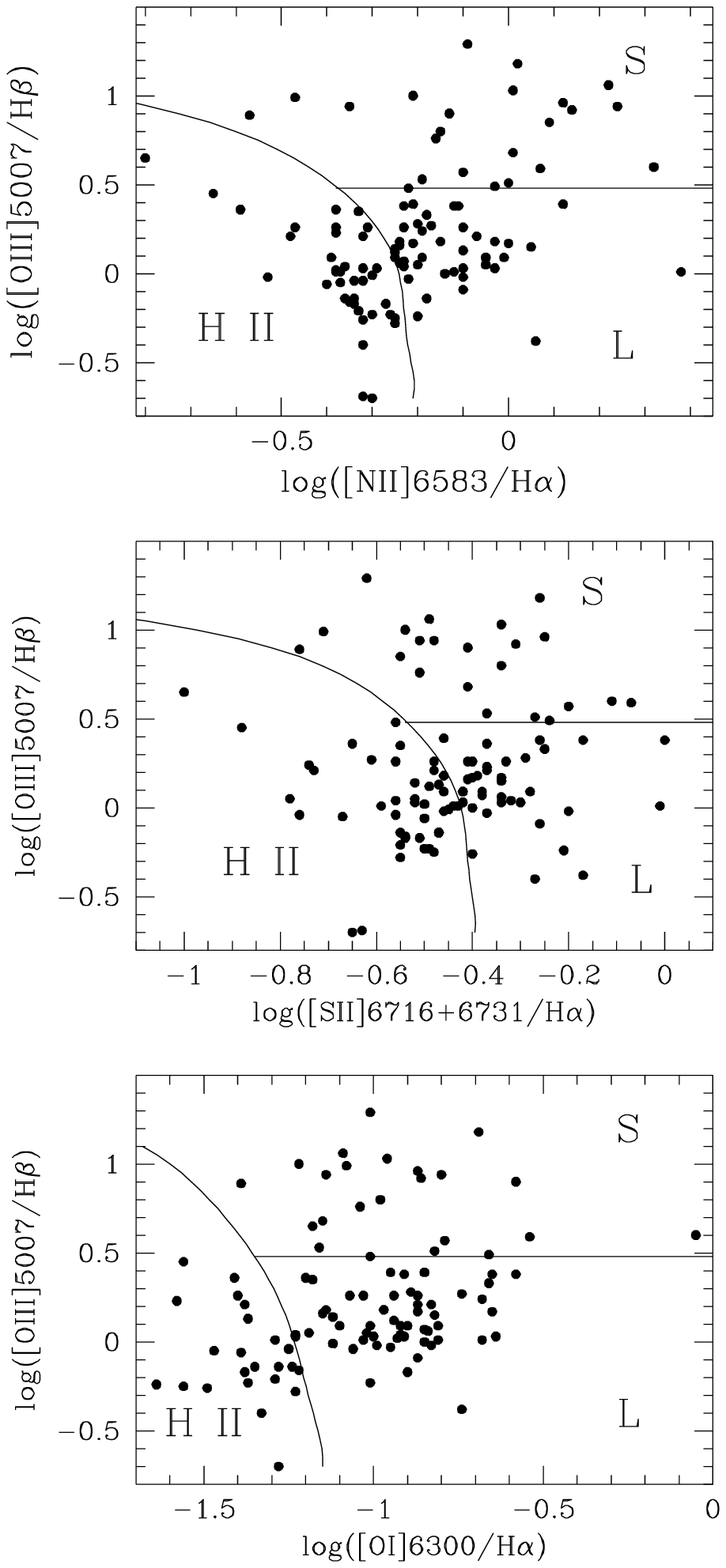]{Dereddened flux ratios.  The H II region-like
galaxies (H) are located to the left of the solid curve while LINERs
(L) and Seyfert 2 galaxies (S) are located on the right.  Seyfert 2
galaxies present [O III]$\lambda$5007/H$\beta\ \geq$ 3 (indicated by a
solid horizontal segment).  Seyfert 1 galaxies are not shown in the
diagrams.  }

\figcaption[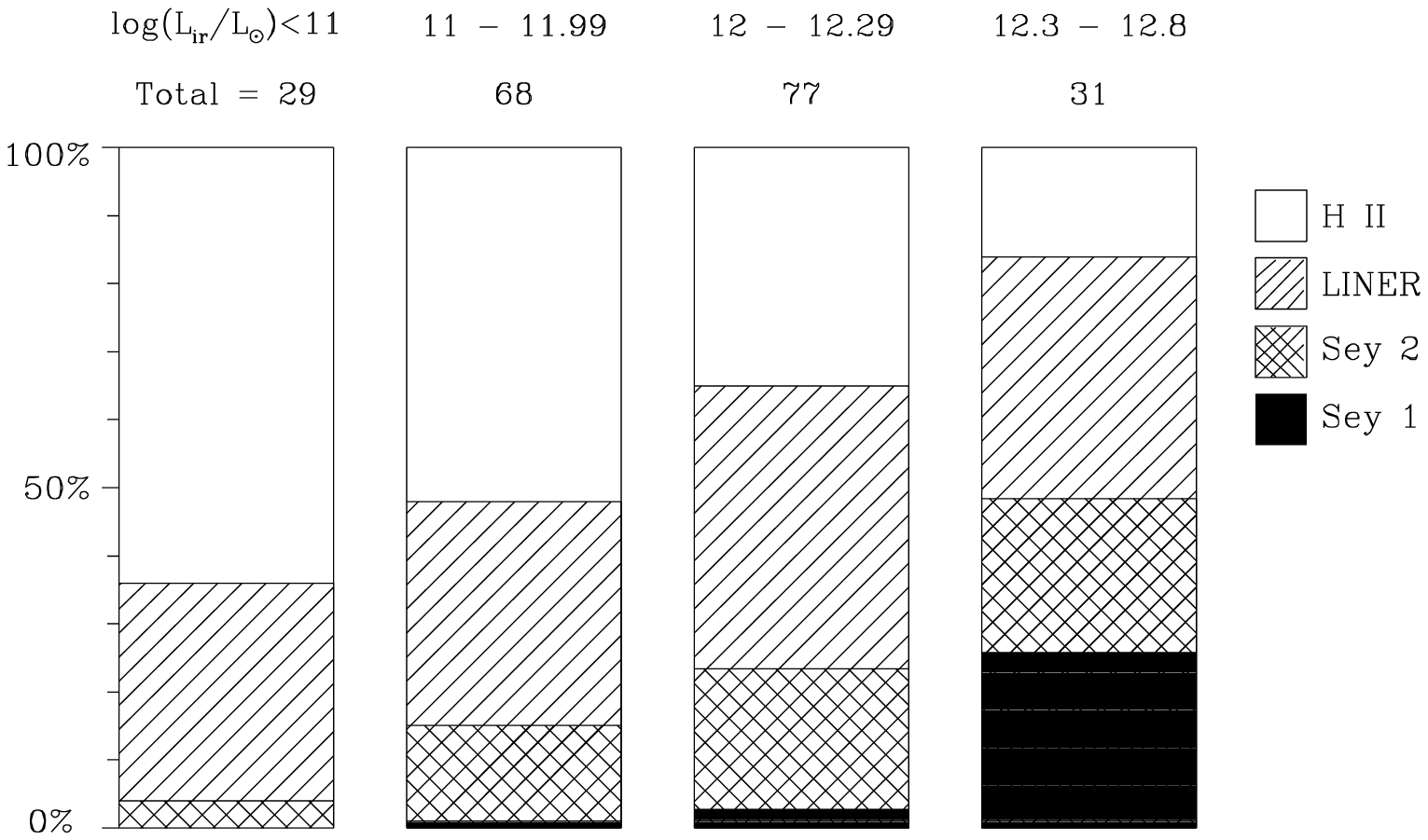]{Summary of the results of the spectral
classification as a function of the infrared luminosity. The results
from the present paper were combined with those obtained by Veilleux
et al. (1995) on LIGs in the BGS.  The fraction of Seyferts increases
with infrared luminosity.  }

\figcaption[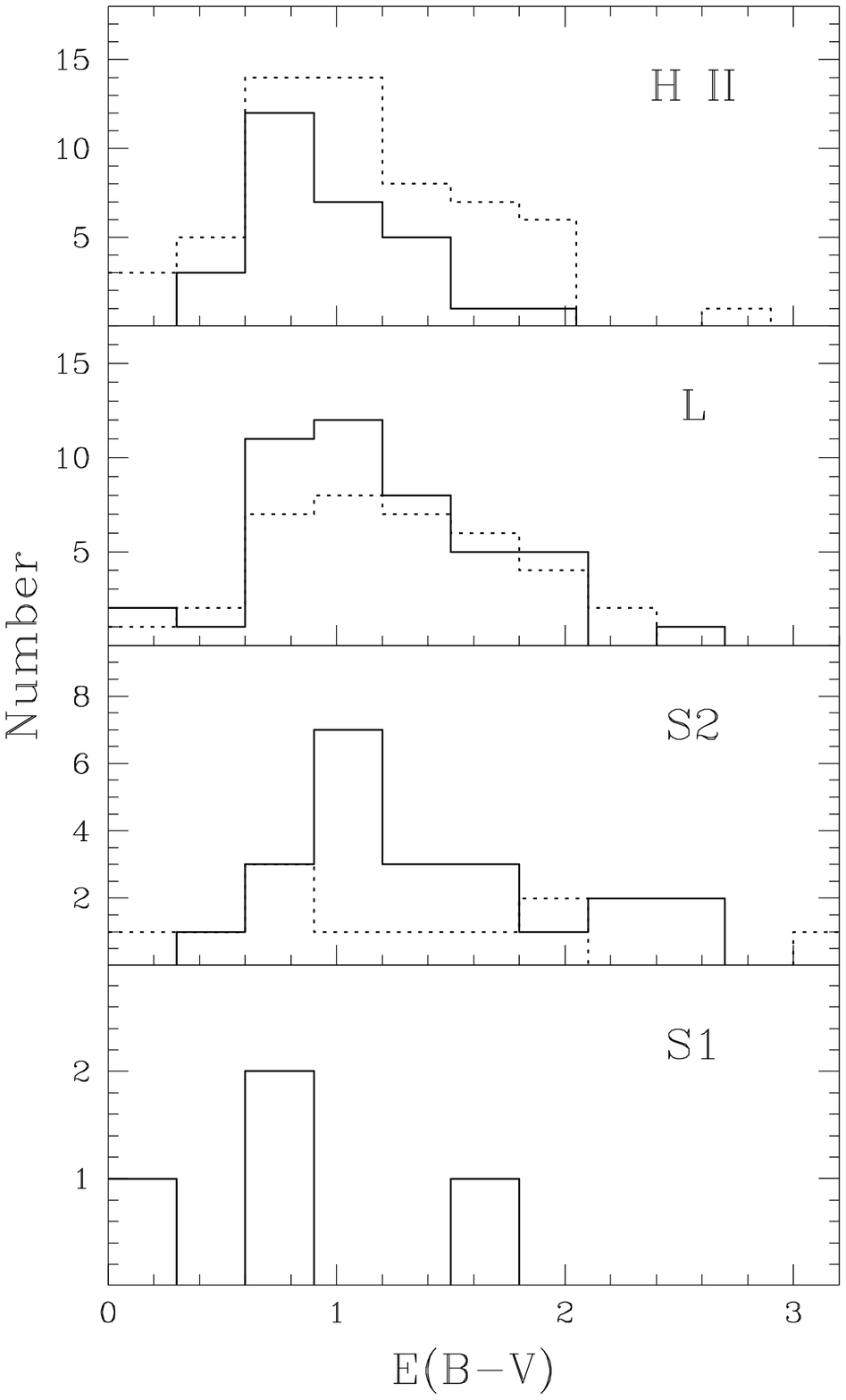]{Distribution of the color excesses as a function
of spectral types for the ULIGs in the 1-Jy sample (solid line) and
the BGS LIGs of Veilleux et al.  (1995; dashed line).  Both classes of
objects present large color excesses.  The differences between the
various types of ULIGs are not significant.  }

\figcaption[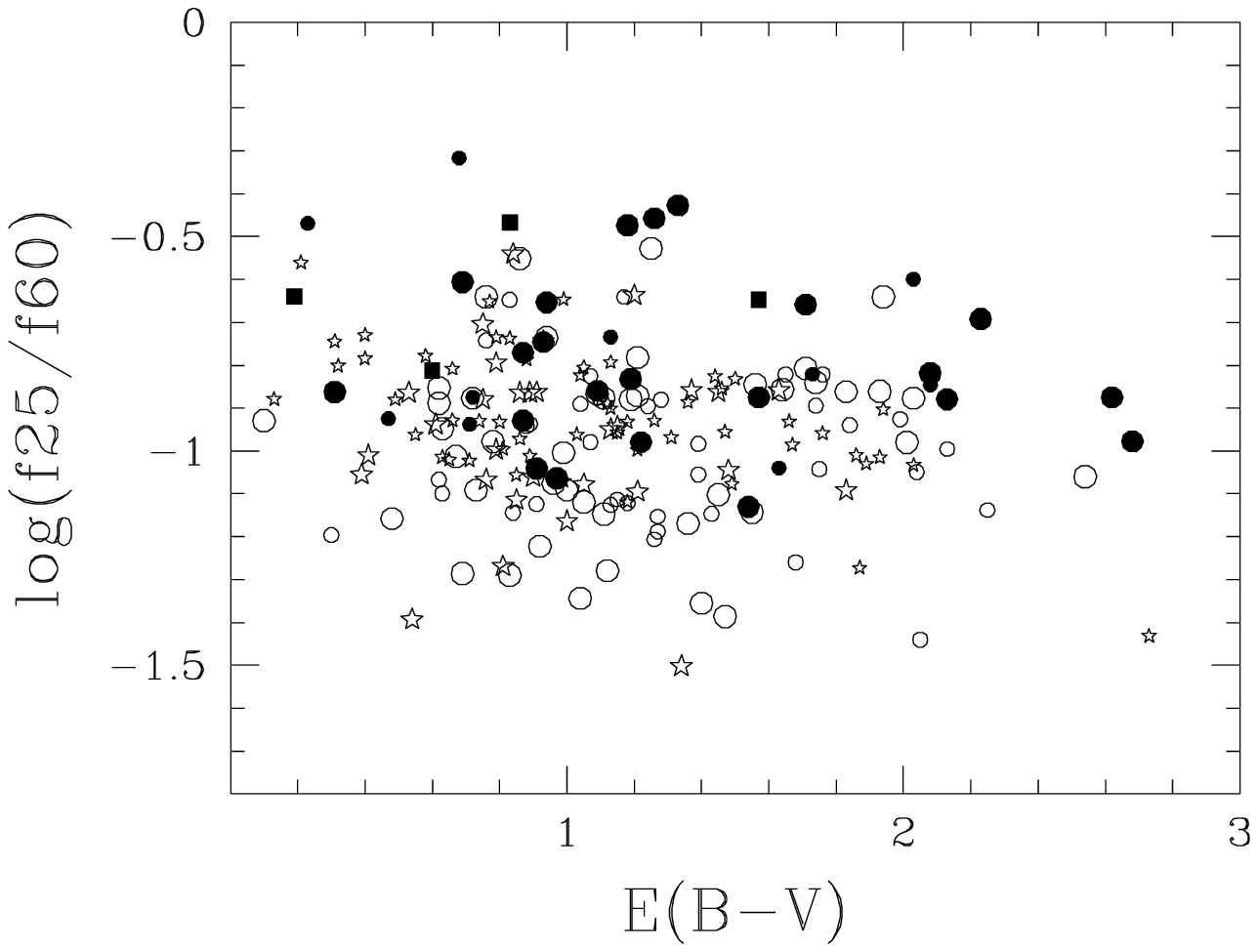]{{\em IRAS} flux density ratios ${\em f}_{25}/{\em
f}_{60}$ as a function of the color excesses for each spectral type.
The stars are the H~II galaxies, the open circles are the LINERs, and
the filled circles are the Seyfert 2 galaxies. The large symbols are
the ULIGs from the 1-Jy sample and the small symbols are the BGS LIGs
from Veilleux et al.  (1995). No significant correlation is observed.
}

\figcaption[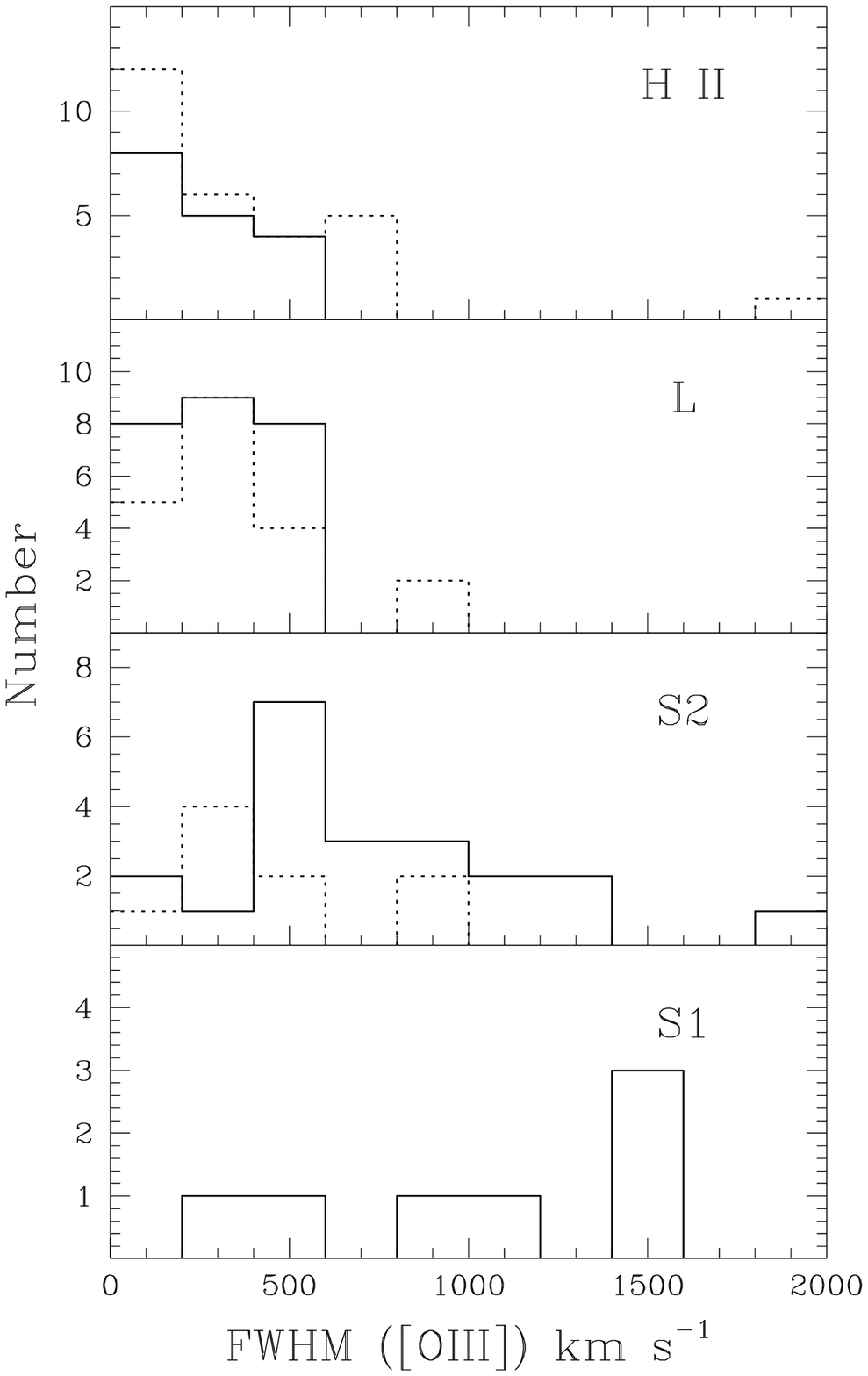]{Distribution of the [O~III] $\lambda$5007 line
widths as a function of the spectral types of the ULIGs in the 1-Jy
sample (solid line) and the BGS LIGs of Veilleux et al.  (1995; dashed
line).  The median line widths of H~II and LINER ULIGs are significantly
smaller than those of Seyfert ULIGs. }

\figcaption[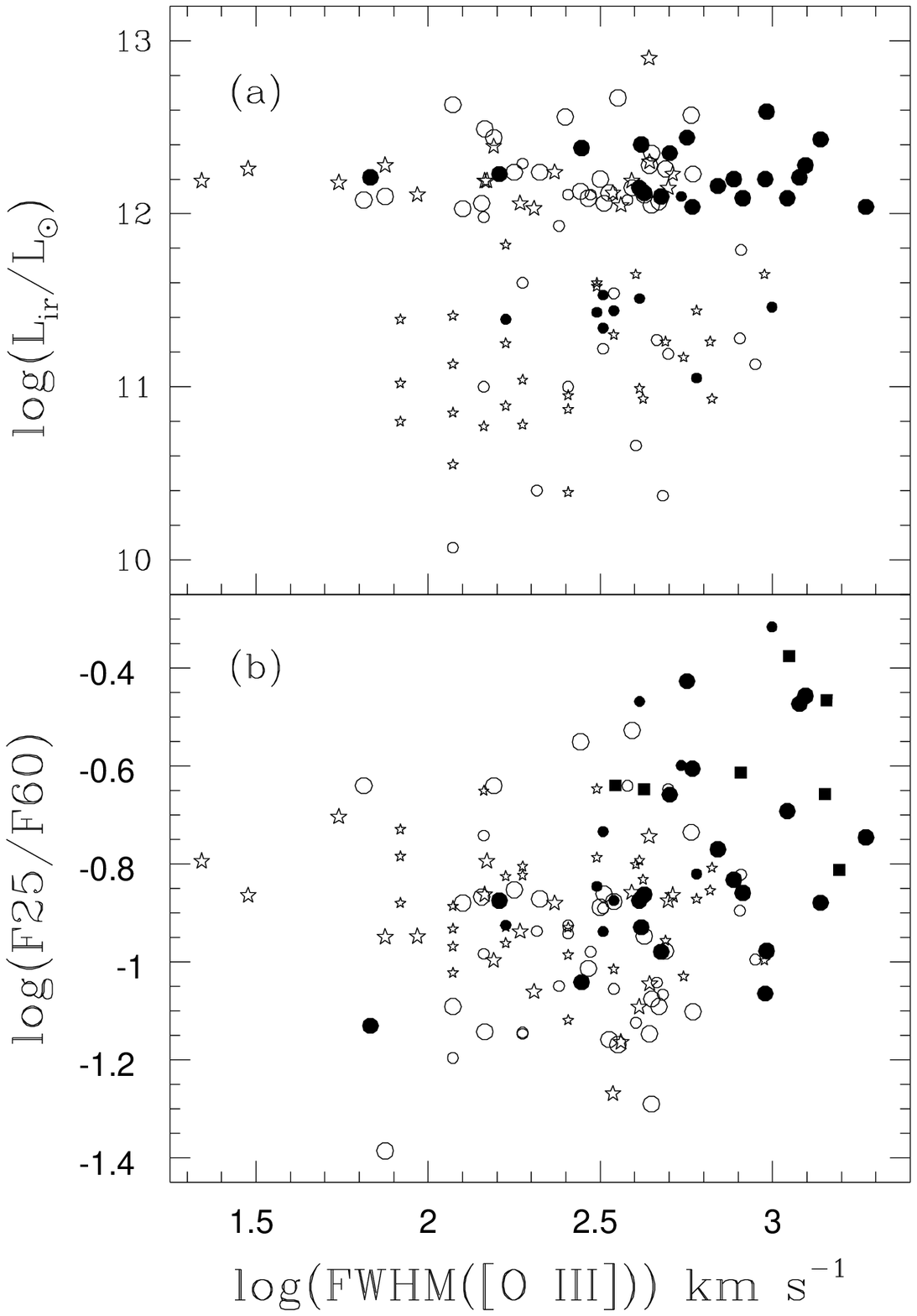]{[O III] $\lambda$5007 line widths as a function
of (a) the infrared luminosities and (b) the {\em IRAS} flux density
ratios ${\em f}_{25}/{\em f}_{60}$. Solid squares represent Seyfert 1
galaxies.  The meaning of the other symbols is the same as in
Fig. 5. The objects with large [O~III] line widths generally have
large infrared luminosities and many of them have Seyfert
optical characteristics and warm {\em IRAS} colors.  }

\figcaption[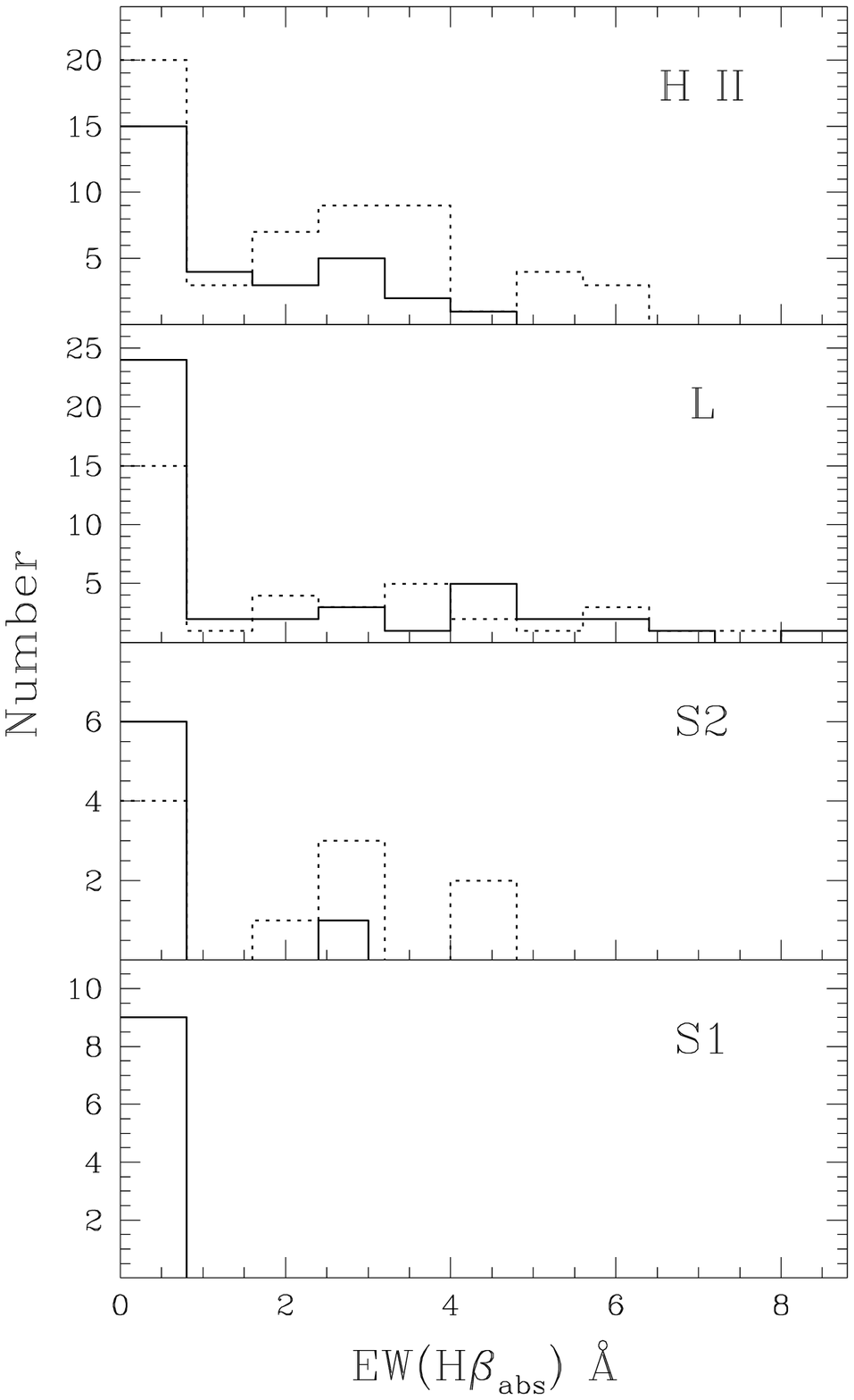]{Distribution of the equivalent widths of
H$\beta$ in absorption as a function of the spectral types of the
ULIGs in the 1-Jy sample (solid line) and the BGS LIGs of Veilleux et
al.  (1995; dashed line).  The equivalent widths of the ULIGs are on
average smaller than those of the lower-luminosity objects. Seyfert
ULIGs also have somewhat smaller equivalent widths than LINERs and
H~II galaxies. }

\figcaption[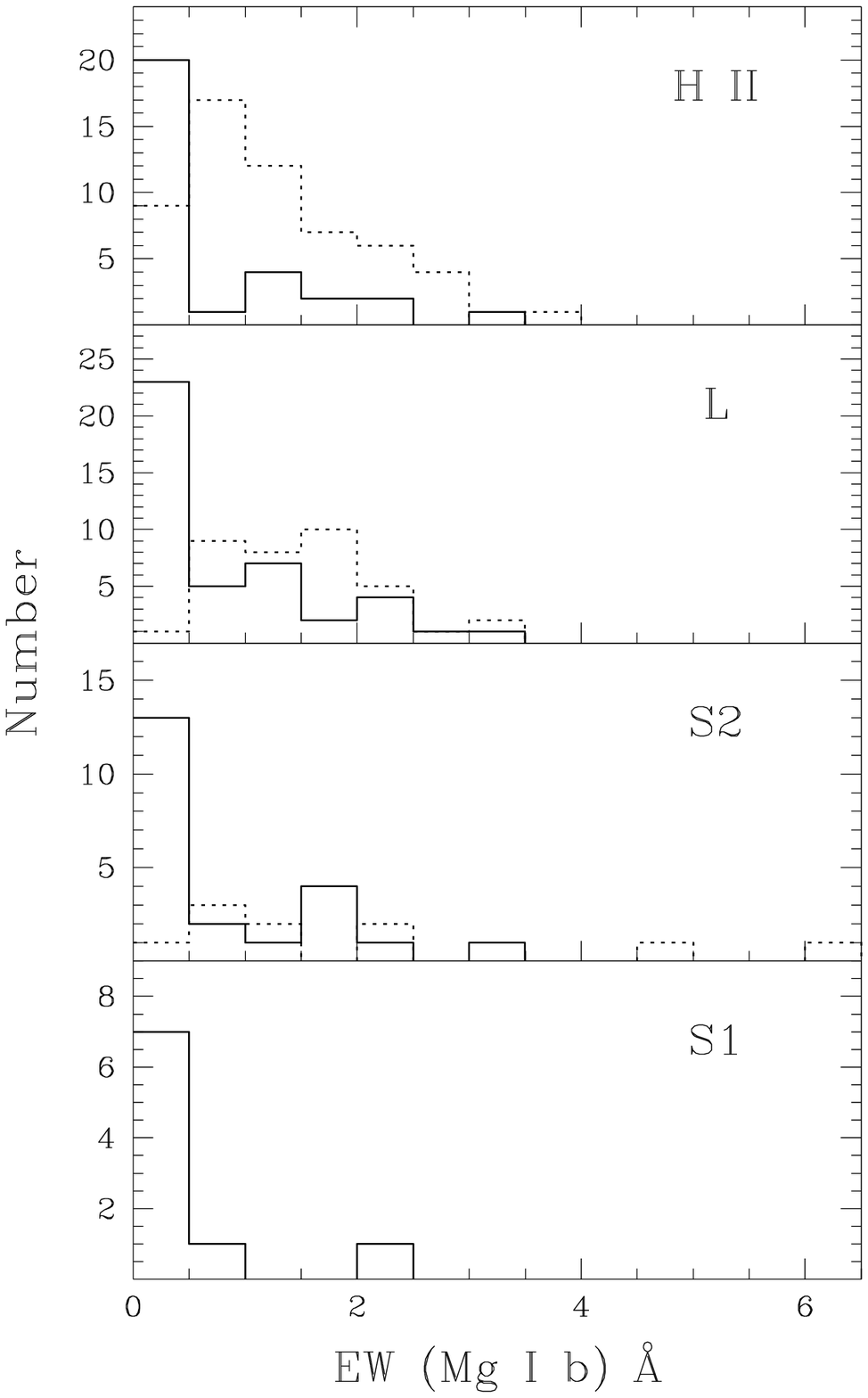]{Distribution of the equivalent widths of Mg~Ib
$\lambda$5176 as a function of the spectral types of the ULIGs in the
1-Jy sample (solid line) and the BGS LIGs of Veilleux et al.  (1995;
dashed line).  The equivalent widths of the ULIGs are on average
smaller than those of the lower-luminosity objects.  Seyfert ULIGs
also have somewhat smaller equivalent widths than LINERs and H~II
galaxies. }

\figcaption[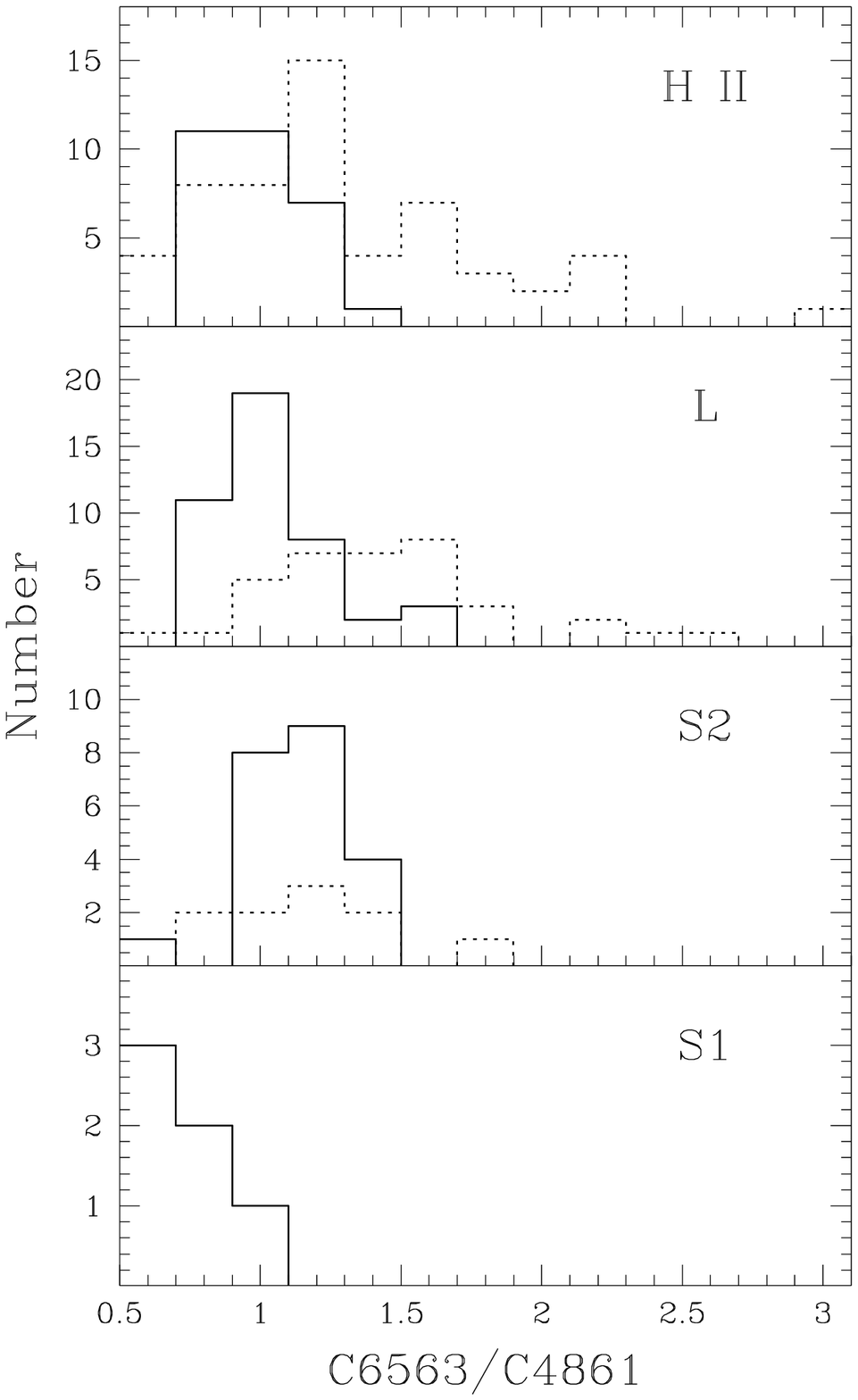]{Distribution of the observed continuum colors,
(C6563/C4861), as a function of the spectral types of the ULIGs in the
1-Jy sample (solid line) and the BGS LIGs of Veilleux et al.  (1995;
dashed line).  The nuclear continuum of Seyfert 1 ULIGs is
significantly bluer on average than the continuum of all other classes
of objects. There is also a tendency for the continuum colors of ULIGs
to be bluer on average than those of LIGs.}

\figcaption[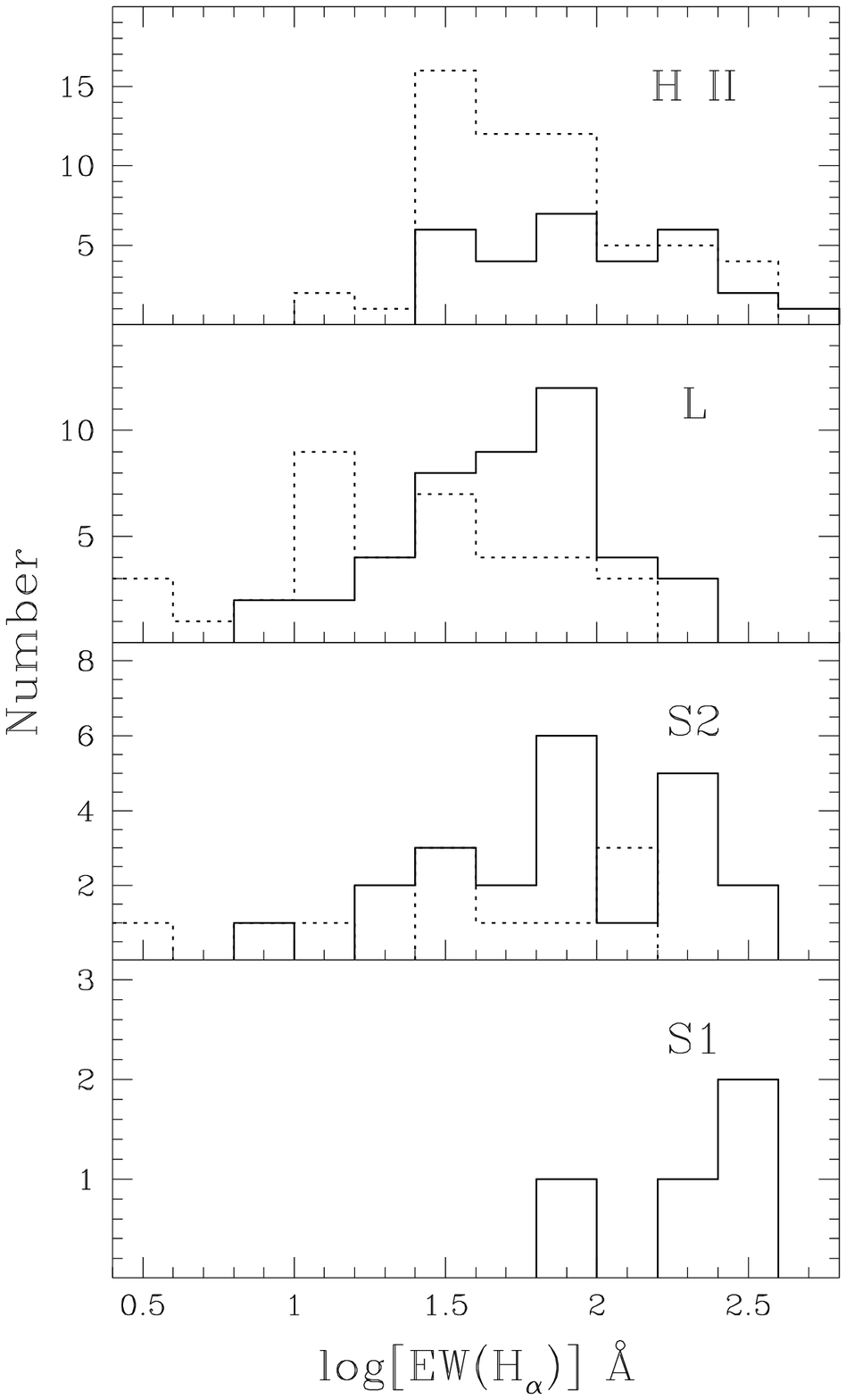]{Distribution of the equivalent widths of
H$\alpha$ in emission as a function of the spectral types of the ULIGs
in the 1-Jy sample (solid line) and the BGS LIGs of Veilleux et al.
(1995; dashed line).  Note the large equivalent widths of Seyfert 1
galaxies.  LINERs have equivalent widths which are somewhat smaller
than those of H~II galaxies, both at low and high infrared luminosities. }

\figcaption[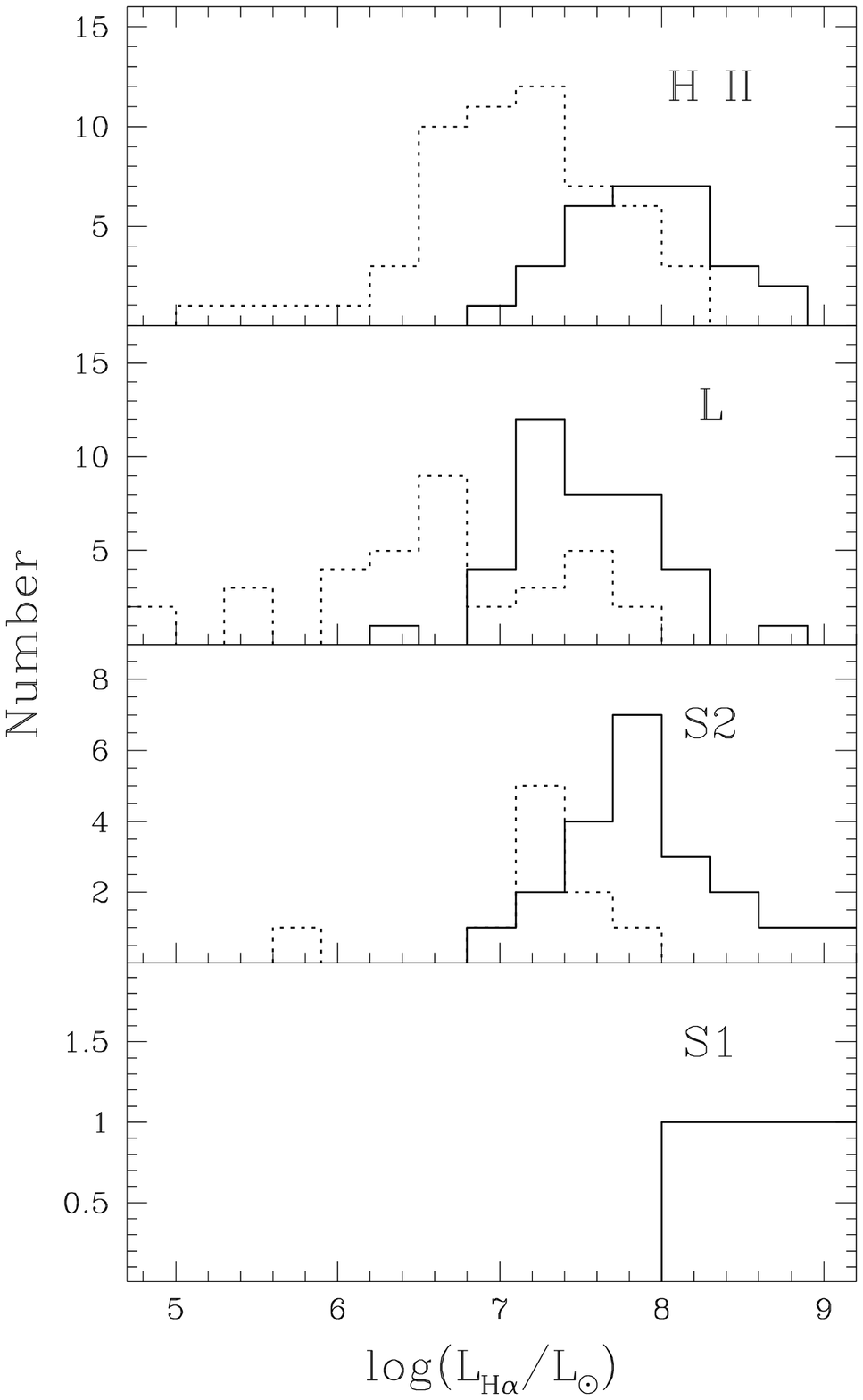]{Distribution of the observed
H$\alpha$ luminosities as a function of the spectral types of the
ULIGs in the 1-Jy sample (solid line) and the BGS LIGs of Veilleux et
al. (1995; dashed line).  Seyfert 1 galaxies are very luminous
H$\alpha$ emitters. The LINERs are somewhat underluminous relative to
H{\ts}II and Seyfert 2 galaxies, both at high and low infrared
luminosities.  }

\figcaption[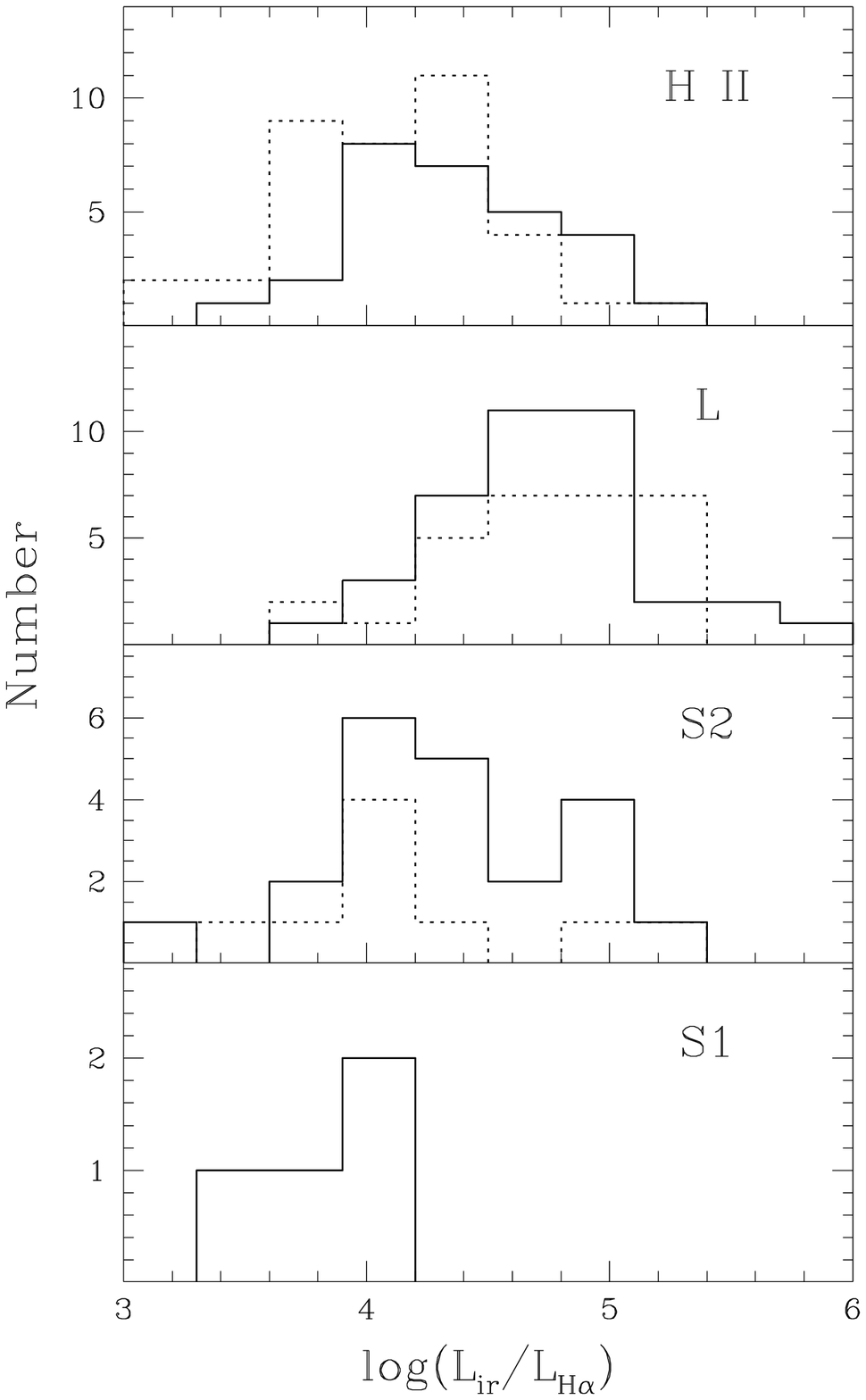]{Distribution of the observed
infrared-to-H$\alpha$ luminosity ratios as a function of spectral
types of the ULIGs in the 1-Jy sample (solid line) and the BGS LIGs of
Veilleux et al. (1995; dashed line).  Seyfert 1s present smaller
ratios than any other classes objects; LINERs present slightly larger
ratios than H~II and Seyfert 2 galaxies. }

\figcaption[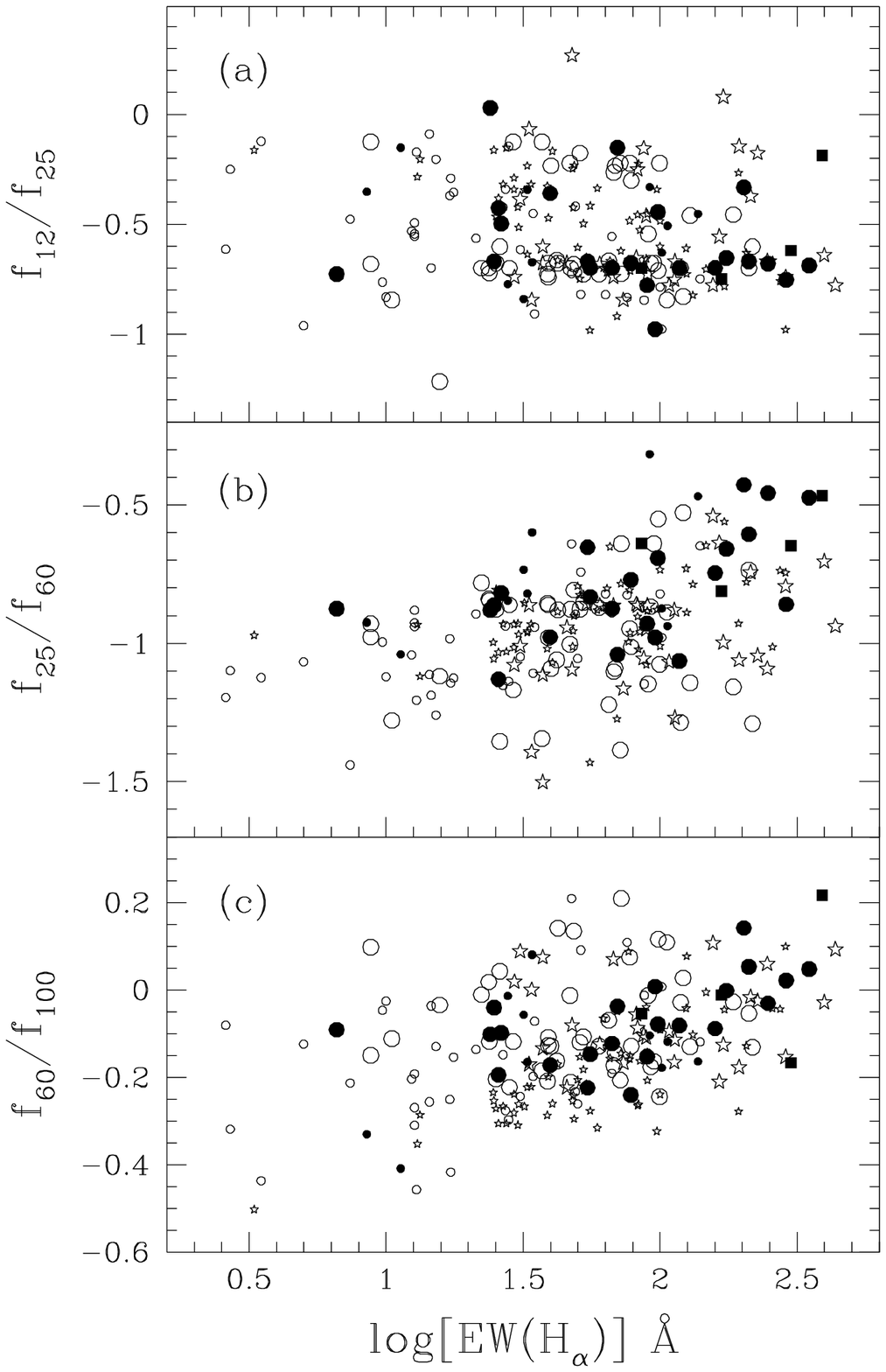]{H$\alpha$ equivalent widths as a function of
(a) {\em IRAS}~flux density ratio ${\em f}_{12}/{\em f}_{25}$, (b)
{\em IRAS} flux density ratio ${\em f}_{25}/{\em f}_{60}$, and (c)
{\em IRAS} flux density ratio ${\em f}_{60}/{\em f}_{100}$. The
meaning of the symbols is the same as in Fig. 7. Correlations are
apparent among H~II galaxies. }

\figcaption[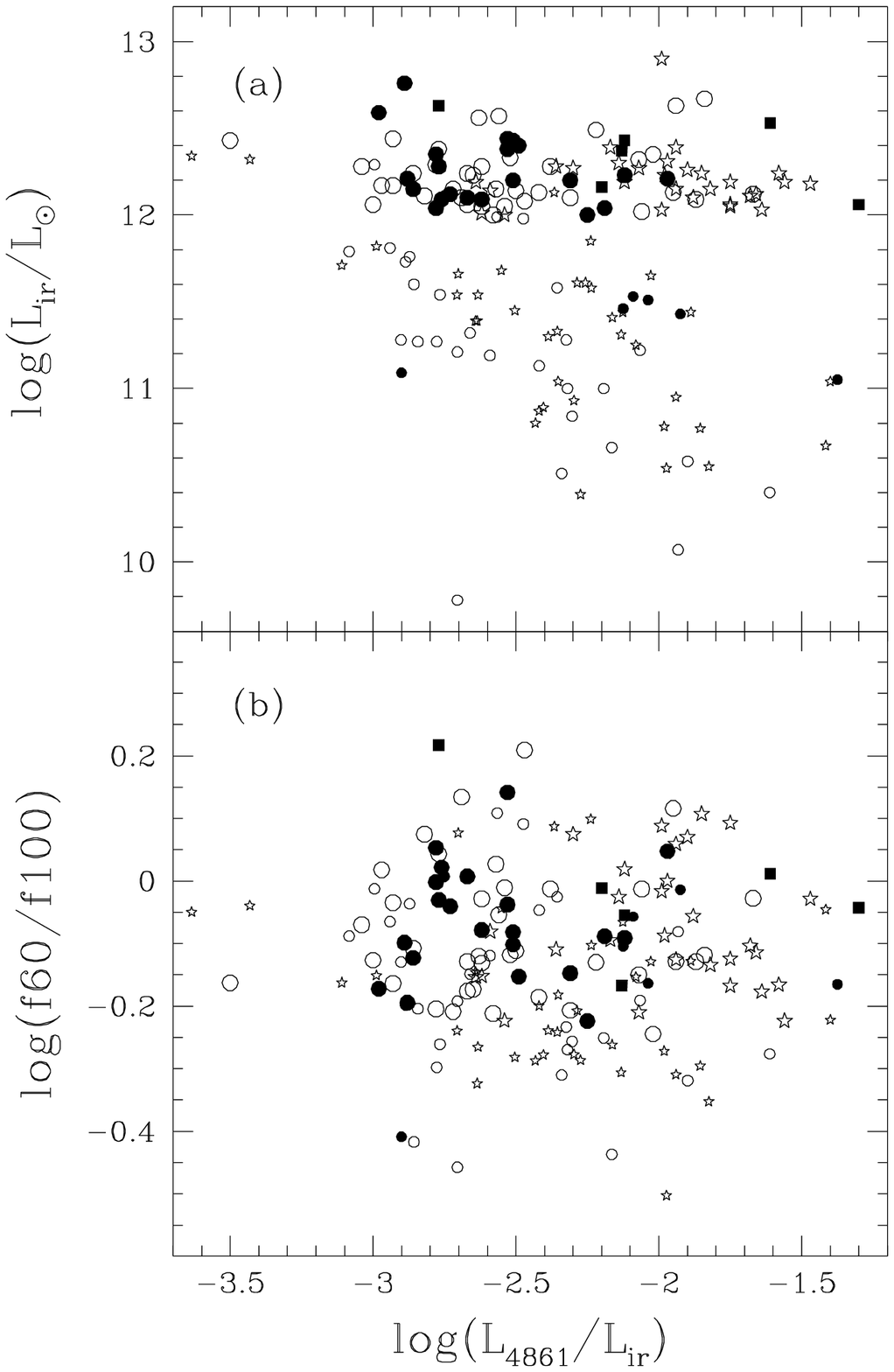]{Observed optical-to-infrared luminosity ratios as a
function of (a) infrared luminosity and (b) {\em IRAS} flux density
ratio ${\em f}_{60}/{\em f}_{100}$. The meaning of the symbols is the
same as in Fig. 7. The correlations previously detected among the BGS
LIGs are not apparent among the 1-Jy ULIGs. }

\figcaption[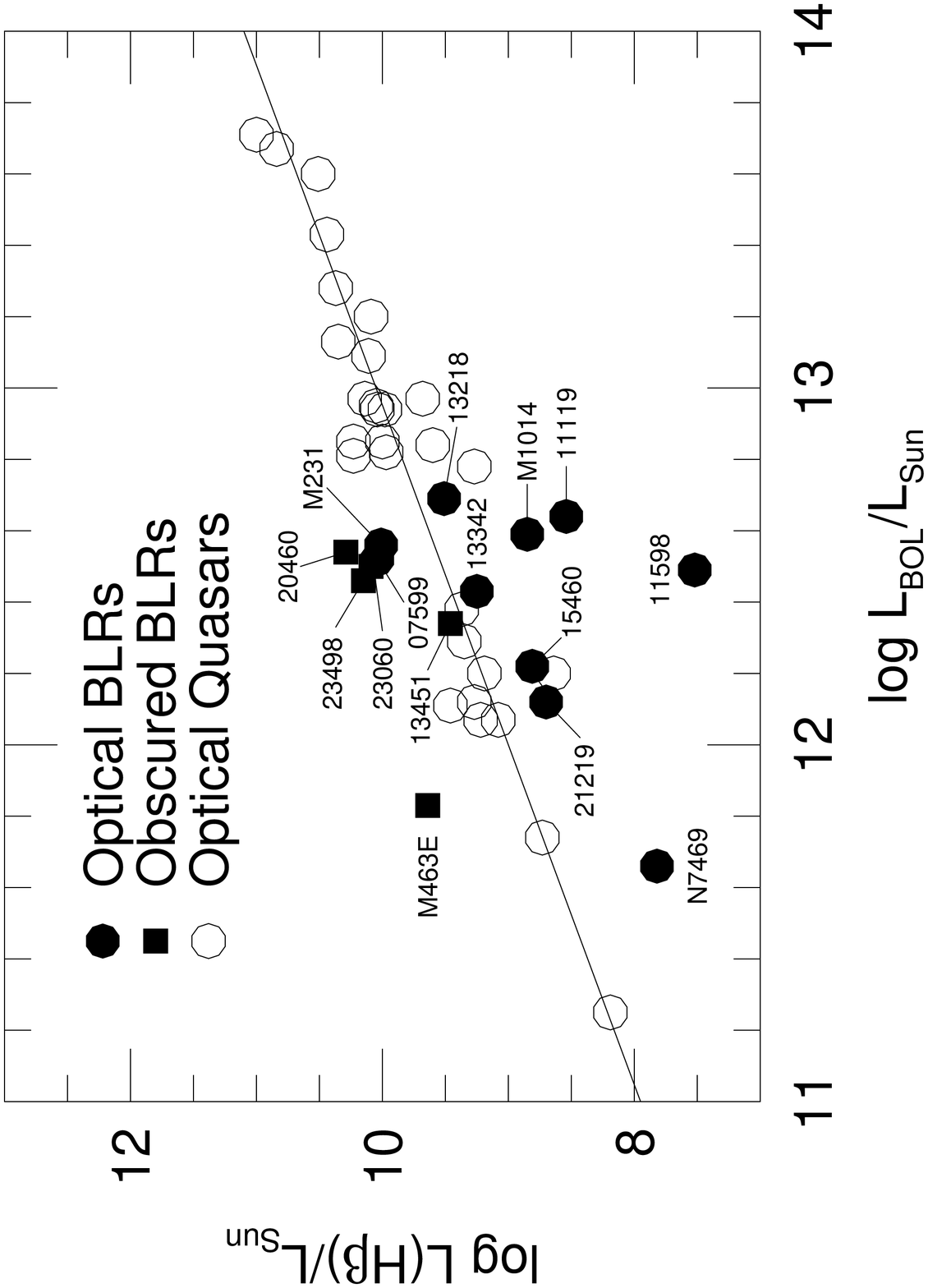]{Comparison of the broad-line H$\beta$
luminosities of ULIGs and optically identified QSOs as a function of
their bolometric luminosities. The data on the optical quasars are
taken from Yee (1980), while the data on ULIGs with obscured
broad-line regions are from Veilleux, Sanders, \& Kim (1997).  See
text for a discussion of the methods used to construct this figure.
The solid line represents the best log-linear fit through the quasar data
points.  Given the uncertainties on the broad-line luminosities (about
30{\ts}\%), only three objects from the sample of ten ULIGs with optical
broad lines (NGC~7469, F11119+3257, and F11598-0112) clearly fall
below the solid line and therefore appear to be powered predominantly
by a starburst. The quasar is likely to be the dominant energy source in
the other ULIGs. }

\clearpage

\setcounter{figure}{0}
\begin{figure}
\plotone{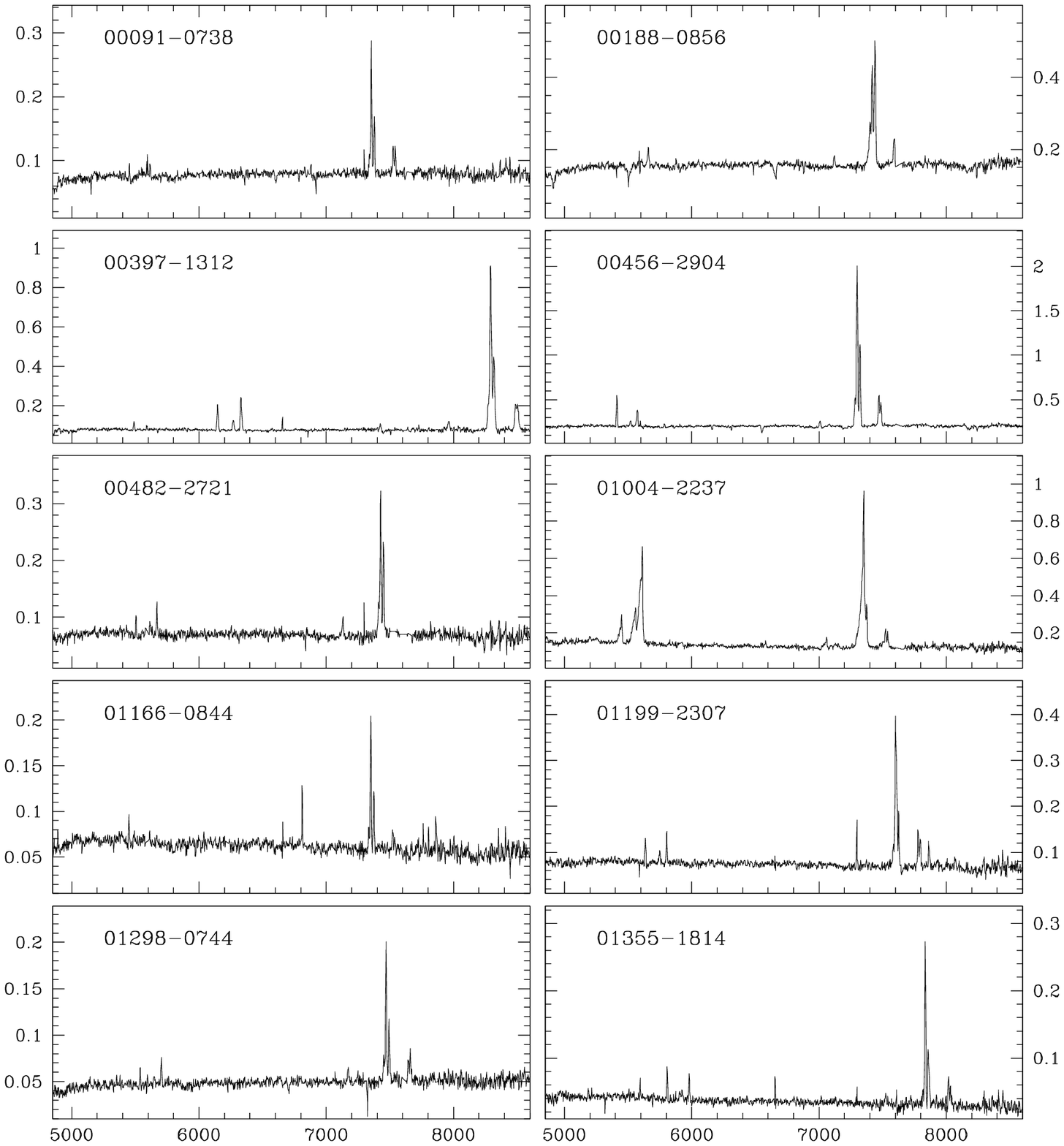}
\caption{ }
\end{figure}

\setcounter{figure}{0}
\begin{figure}
\plotone{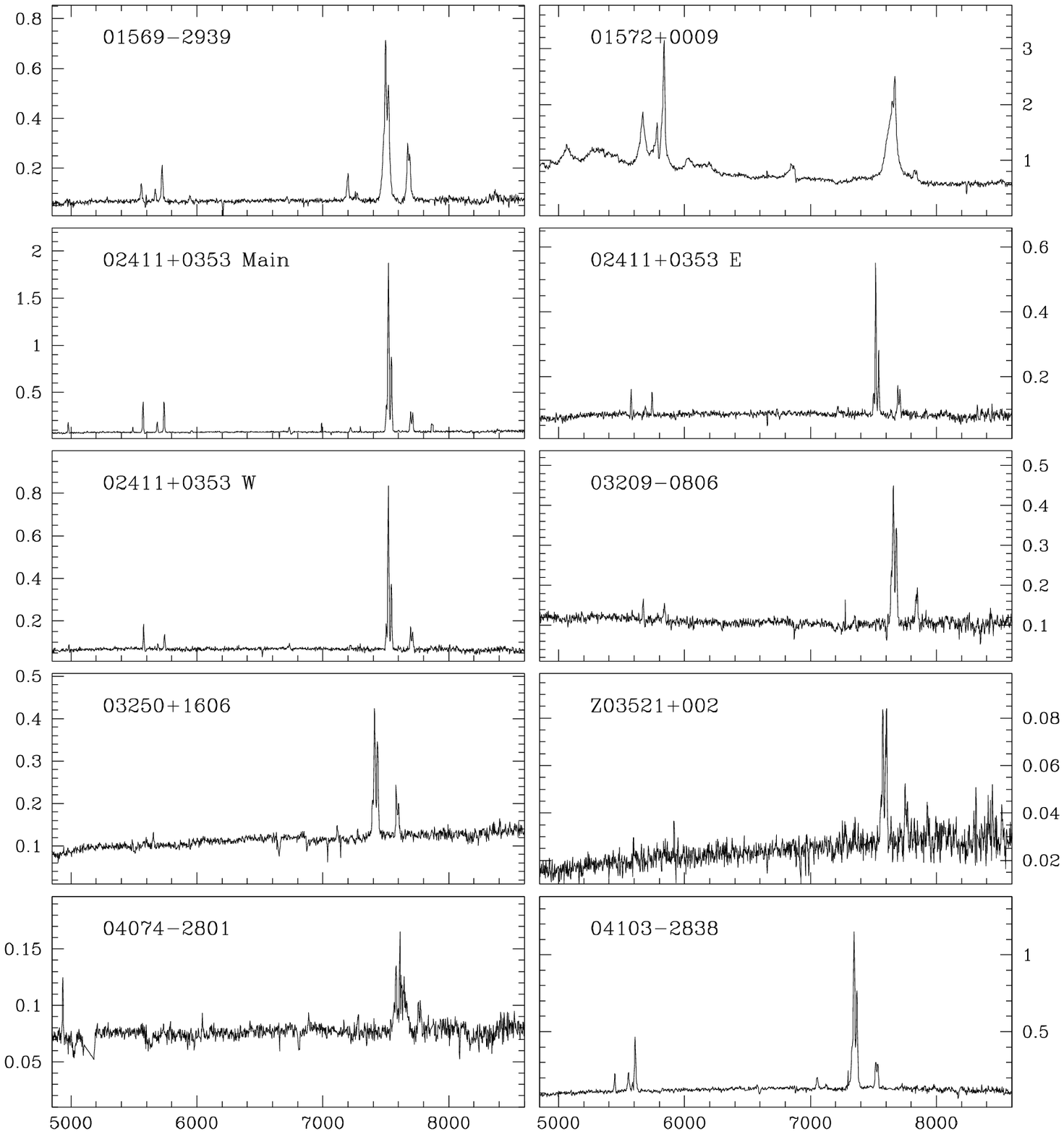}
\caption{ }
\end{figure}

\setcounter{figure}{0}
\begin{figure}
\plotone{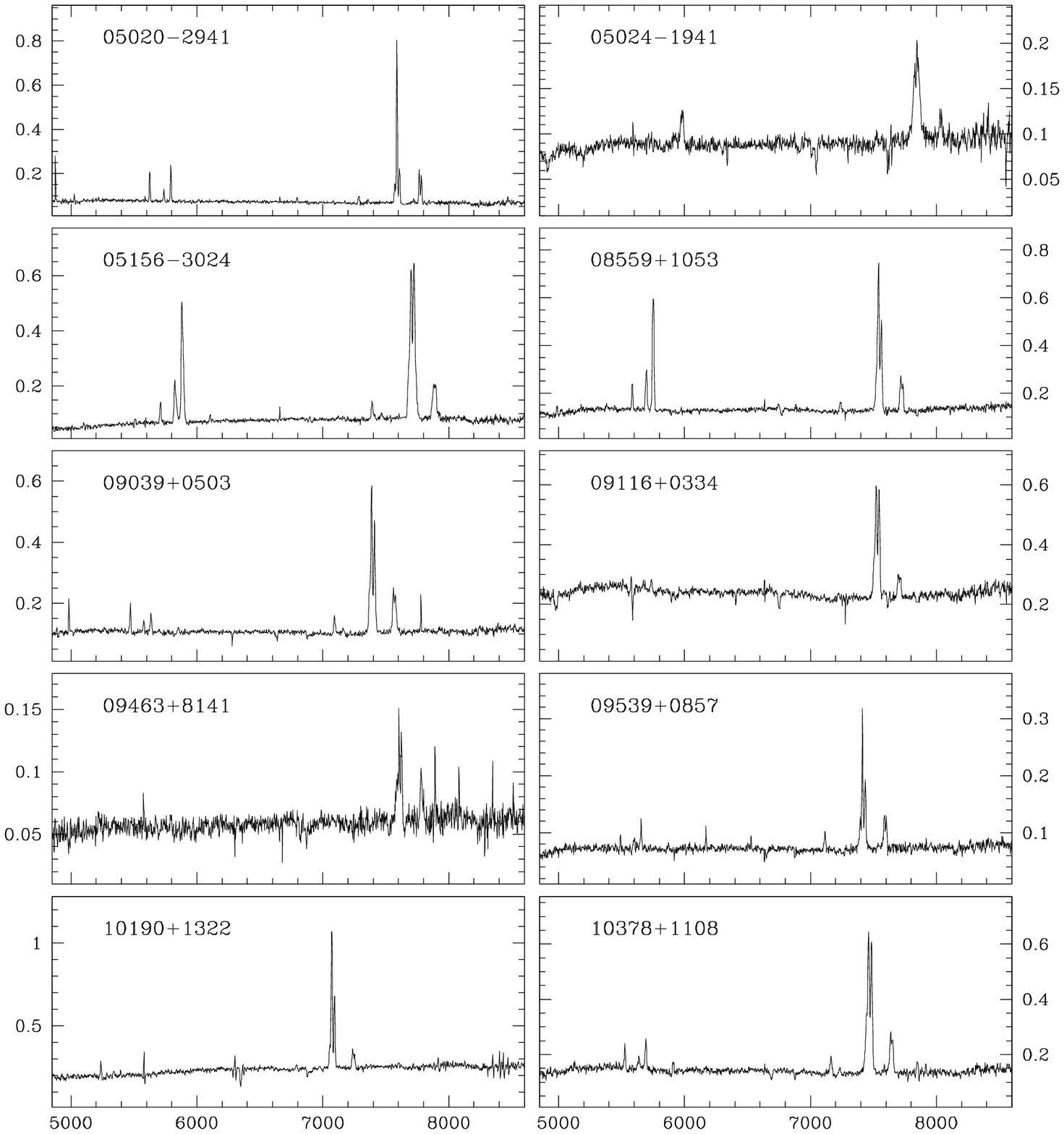}
\caption{ }
\end{figure}

\setcounter{figure}{0}
\begin{figure}
\plotone{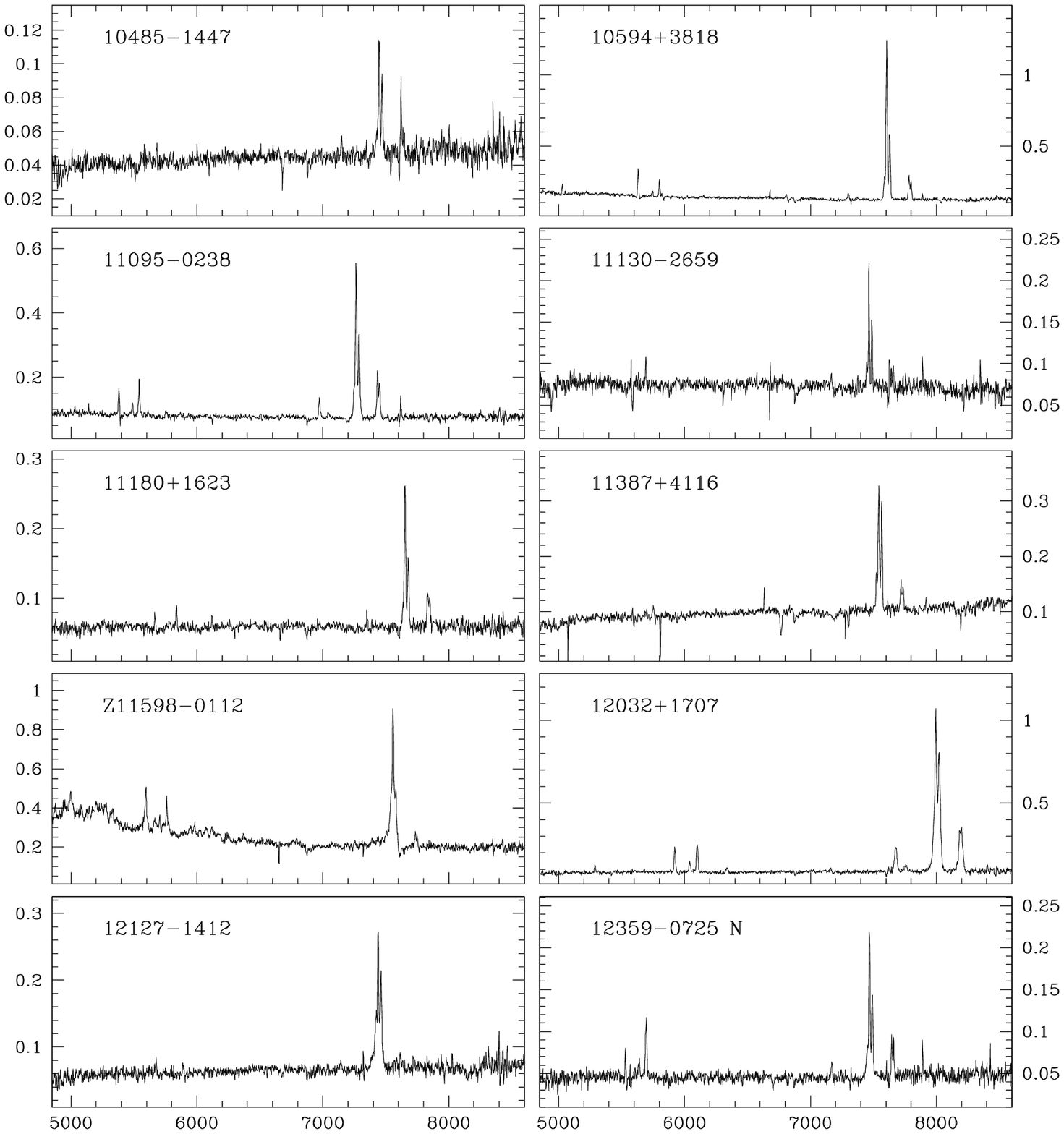}
\caption{ }
\end{figure}

\setcounter{figure}{0}
\begin{figure}
\plotone{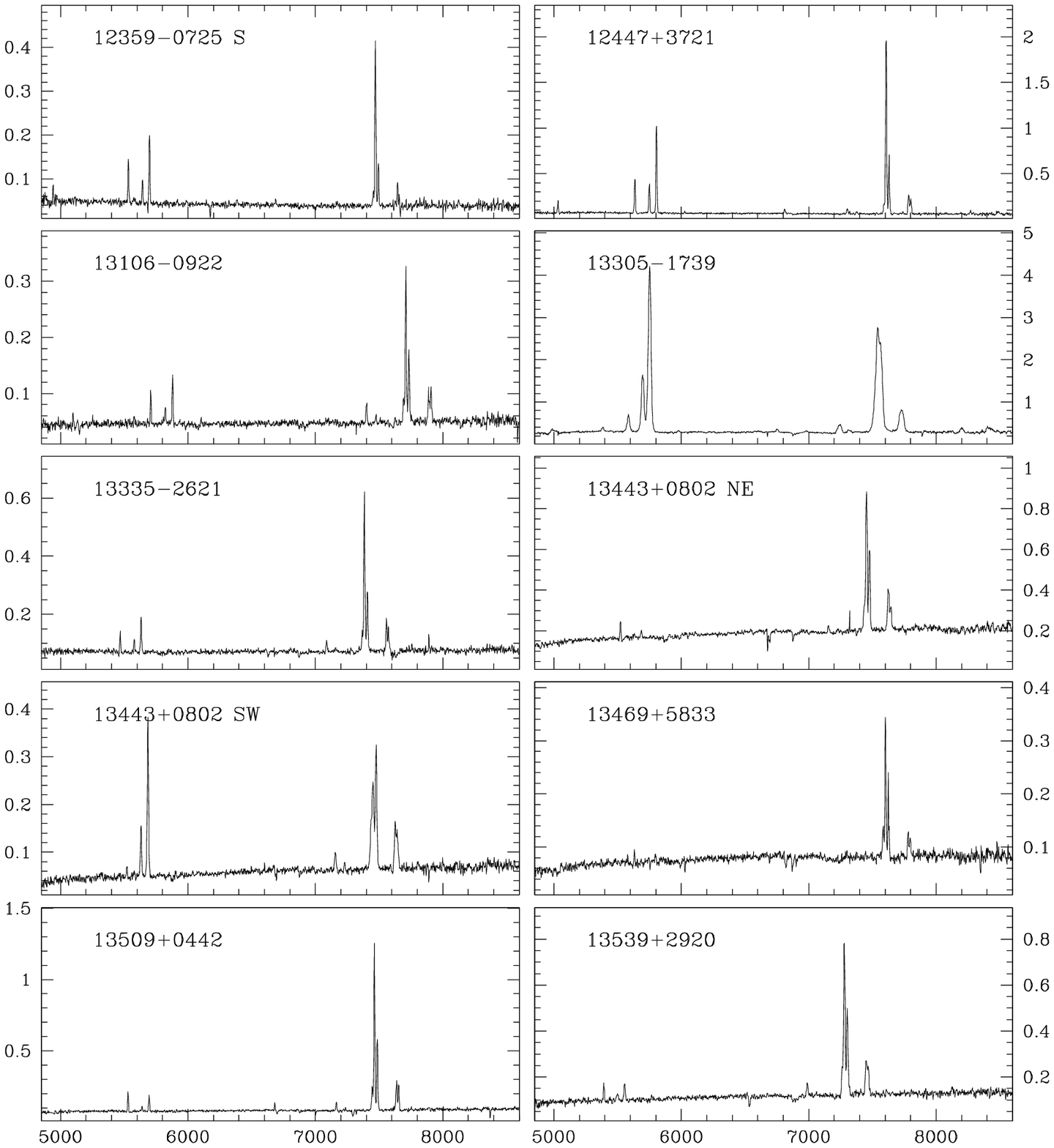}
\caption{ }
\end{figure}

\setcounter{figure}{0}
\begin{figure}
\plotone{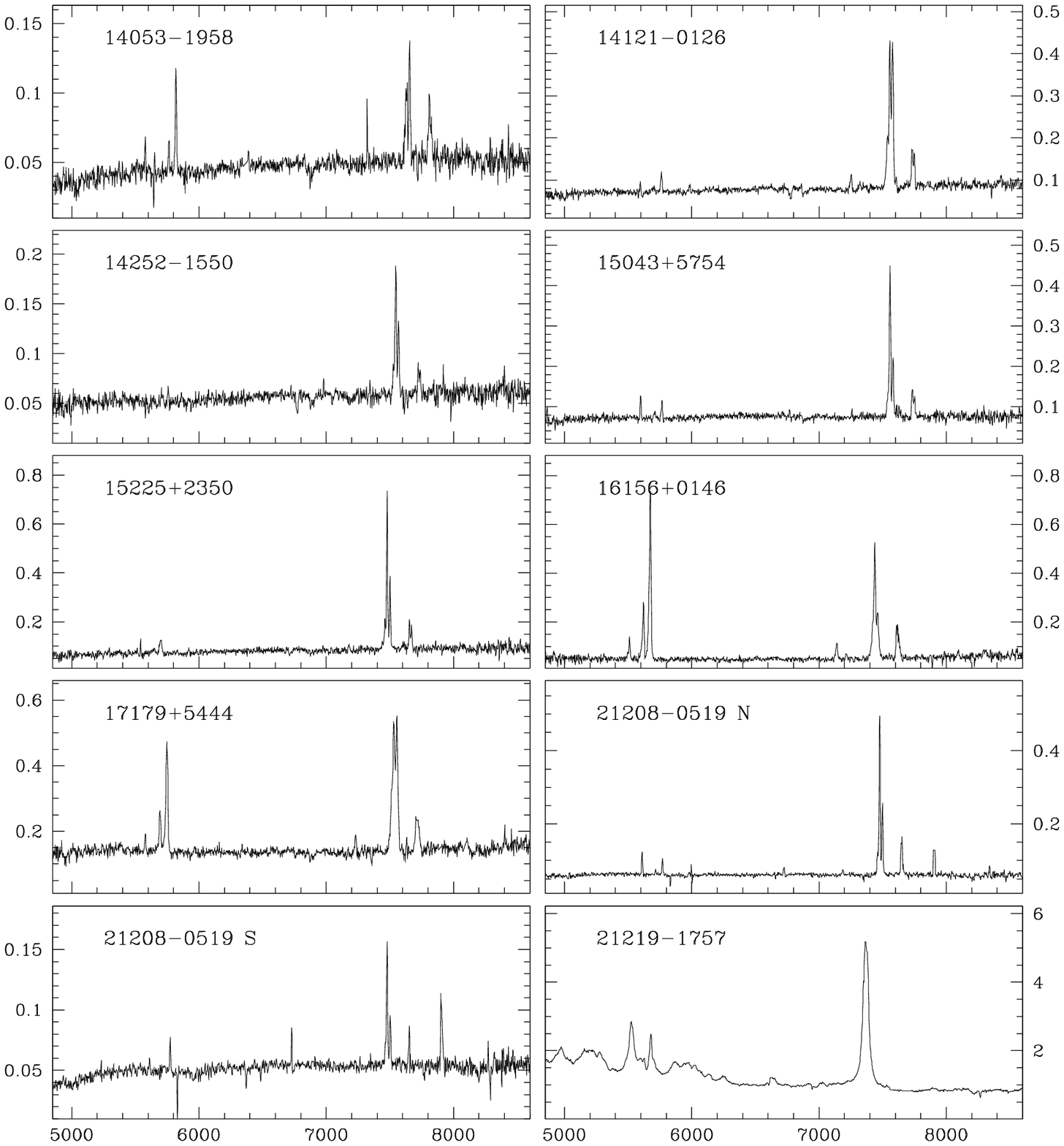}
\caption{ }
\end{figure}

\setcounter{figure}{0}
\begin{figure}
\plotone{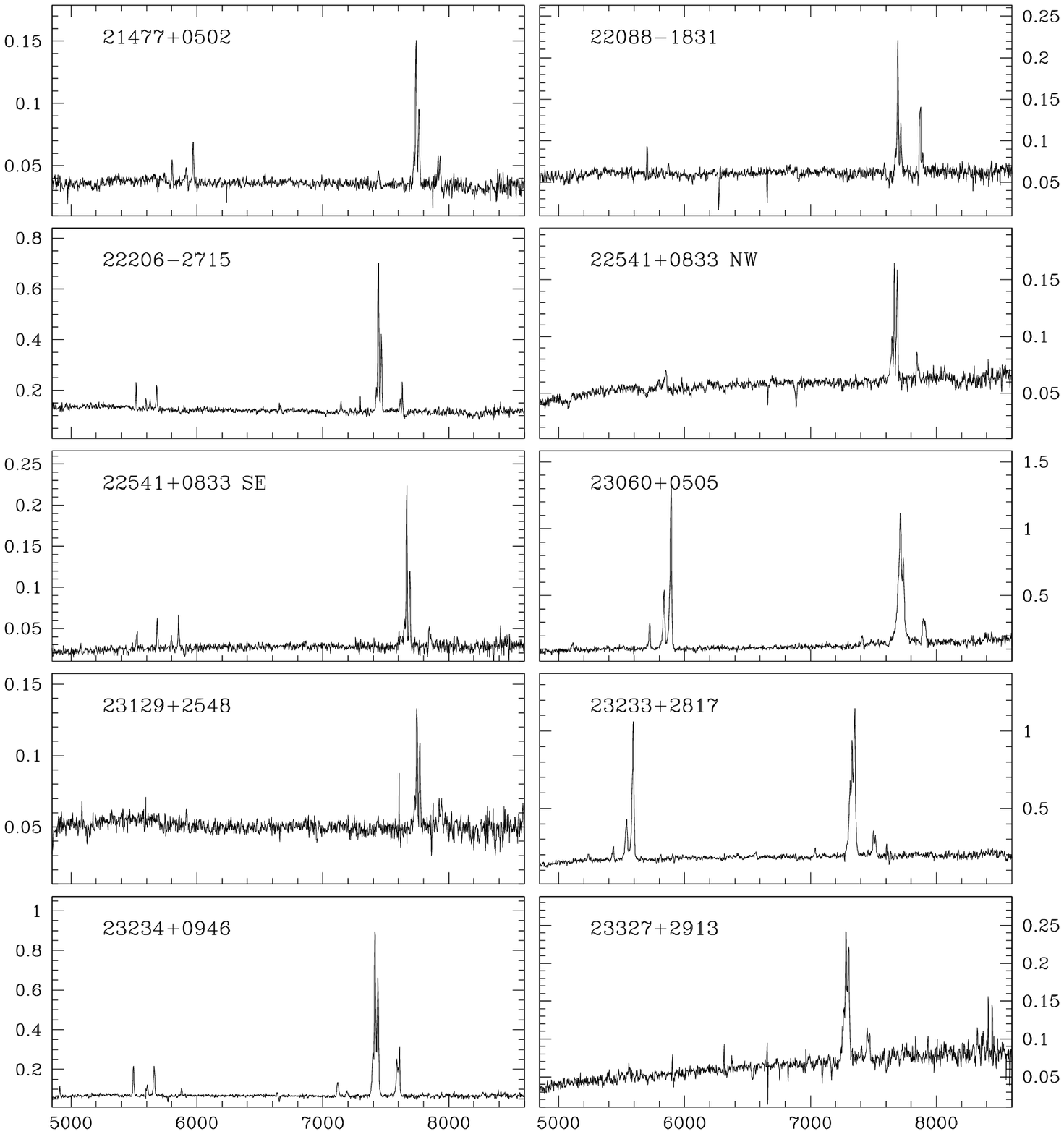}
\caption{ }
\end{figure}

\setcounter{figure}{0}
\begin{figure}
\plotone{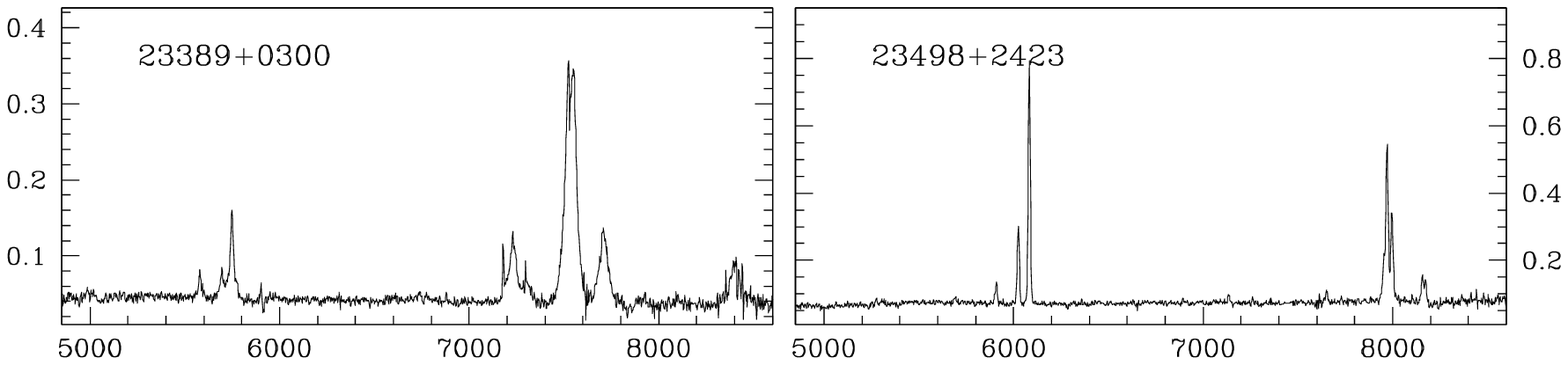}
\caption{ }
\end{figure}

\setcounter{figure}{1}
\begin{figure}
\plotone{fig2.eps}
\caption{ }
\end{figure}

\begin{figure}
\plotone{fig3.eps}
\caption{ }
\end{figure}

\begin{figure}
\plotone{fig4.eps}
\caption{ }
\end{figure}

\begin{figure}
\plotone{fig5.eps}
\caption{ }
\end{figure}

\begin{figure}
\plotone{fig6.eps}
\caption{ }
\end{figure}

\begin{figure}
\plotone{fig7.eps}
\caption{ }
\end{figure}

\begin{figure}
\plotone{fig8.eps}
\caption{ }
\end{figure}

\begin{figure}
\plotone{fig9.eps}
\caption{ }
\end{figure}

\begin{figure}
\plotone{fig10.eps}
\caption{ }
\end{figure}

\begin{figure}
\plotone{fig11.eps}
\caption{ }
\end{figure}

\clearpage

\begin{figure}
\plotone{fig12.eps}
\caption{ }
\end{figure}

\begin{figure}
\plotone{fig13.eps}
\caption{ }
\end{figure}

\begin{figure}
\plotone{fig14.eps}
\caption{ }
\end{figure}

\begin{figure}
\plotone{fig15.eps}
\caption{ }
\end{figure}

\begin{figure}
\plotone{fig16.eps}
\caption{ }
\end{figure}

\end{document}


\newcommand{\ts}{\thinspace}
\begin{deluxetable}{lccccc}
\tablenum{1}
\tablewidth{0pt}
\tablecaption{Journal of Observations}
\tablehead{
\colhead{Obs. Date} &
\multicolumn{1}{c}{Grating} &
\multicolumn{1}{c}{Spectral Range} &
\multicolumn{1}{c}{Resolution} &
\multicolumn{1}{c}{Seeing} &
\colhead{CCD} \nl
\colhead{} &
\multicolumn{1}{c}{(l/mm)} &
\multicolumn{1}{c}{(\AA)} &
\multicolumn{1}{c}{(\AA)} &
\multicolumn{1}{c}{($^{\prime\prime}$)} & 
\colhead{} 
}
\startdata
1994 Feb 2-5&300&4550 $-$ 8940&8.3&1.5--3.0&Ford F3KA 3072x512\nl
1995 Mar 26-29&300&4720 $-$ 8990&8.3&2.1--3.0&Ford F3KC 3072x512 \nl
1995 Oct 24-28&300&4080 $-$ 8970&8.3&1.0--1.5&Ford F3KC 3072x512 \nl
\enddata
\end{deluxetable}

\clearpage

\begin{deluxetable}{lcccc}
\tablenum{6}
\tablewidth{4.5in}
\tablecaption{Spectral Classification as a Function of {\it L}$_{\rm ir}$}
\tablehead{
\colhead{log $\left( {L_{\rm ir} \over L_\odot} \right)$ } &
\multicolumn{1}{c}{$<$11\tablenotemark{a}} &
\multicolumn{1}{c}{11--11.99\tablenotemark{a}} &
\multicolumn{1}{c}{12--12.29} &
\multicolumn{1}{c}{12.3--12.8} \nl
}
\startdata
 & & \nl
Number&29&68&77&31 \nl
 & & \nl
H II region (\%)&62&53&35&16 \nl
LINER     (\%)&31&32&41&35 \nl
Seyfert 2 (\%)&7&13&21&23 \nl
Seyfert 1 (\%)&0&2&3&26 \nl
\tablenotetext{a}{Data from luminous infrared galaxies in the BGS studied by VKSMS.}
\enddata
\end{deluxetable}